\documentclass[twocolumn,prl,english]{revtex4-1}
\usepackage[T1]{fontenc,xcolor}
\usepackage[latin9]{inputenc}
\usepackage{verbatim}
\usepackage{amsmath,bm}
\usepackage{amssymb}
\usepackage{graphicx}
\usepackage{babel}
\usepackage{subfigure}

\makeatletter

\newcommand{\e}{\mathrm{e}}
\newcommand{\vecr}{\mathbf{r}}
\newcommand{\vecl}{{\bm \ell}}
\newcommand{\vecp}{\mathbf{p}}

\newcommand{\vecv}{\mathbf{v}}

\newcommand{\vecg}{\mathbf{g}}
\newcommand{\omegav}{\bm \omega}
\newcommand{\Omegav}{\bm \Omega}
\newcommand{\tauv}{\bm \tau}
\newcommand{\ellv}{\bm \ell}

\newcommand{\D}{\mathrm{d}}
\newcommand{\half}{\frac{1}{2}}
\newcommand{\tell}{\tilde{\ell}}
\newcommand{\Drho}{\delta\rho}

\makeatother

\begin{document}

\title{Odd viscosity in active matter: microscopic origin and 3D effects}

\author{Tomer Markovich$^{1}$}
\email{tm36@rice.edu} 
\author{Tom C. Lubensky$^{2}$}
\affiliation{
$^{1}$Center for Theoretical Biological Physics, Rice University, Houston, TX 77005, USA \\
$^{2}$Department of Physics and Astronomy, University of Pennsylvania, Philadelphia, Pennsylvania 19104, USA
}

\date{\today}

\begin{abstract}
In common fluids, viscosity is associated with dissipation. However, when time-reversal-symmetry  is broken a new type of non-dissipative `viscosity' emerges.
Recent theories and experiments on classical 2D systems with active spinning particles have heightened interest in odd viscosity,
but a microscopic theory for it in active materials is still absent.
Here we present such first-principles microscopic Hamiltonian theory, valid for both 2D and 3D, 
showing that odd viscosity is present in any system, even at zero temperature, with globally or locally aligned spinning components.
Our work substantially extends the applicability of odd viscosity into 3D fluids, and specifically to internally driven active materials, such as living matter ({\it e.g.}, actomyosin gels).
We find intriguing 3D effects of odd viscosity such as propagation of anisotropic bulk shear waves and breakdown of Bernoulli's principle.
\end{abstract}
\maketitle


Active materials are composed of many components that convert energy 
from their environment into directed mechanical motion.
Time reversal symmetry (TRS) is thus locally broken leading to novel phenomena such as motility-induced phase separation~\cite{MIPS}, 
giant density fluctuations~\cite{Ramaswamy2005,Marchetti2013}, and entropy production in the (non-equilibrium) steady state~\cite{EPR,Nardini2017}.
Examples of active matter are abundant and range from living matter such as bacteria~\cite{Ariel2019,Aranson2012}, cells~\cite{Tissue2019,fred2010},
actomyosin networks~\cite{Marchetti2013,Joanny2015,PRL2019}, 
and bird flocks~\cite{Flock} to driven Janus particles~\cite{light2019,Spec2013}, colloidal rollers~\cite{Bartolo2015,Bartolo2013}, and macroscale driven chiral rods~\cite{Tsai2005}. 

One of the striking phenomena arising from broken TRS is the possible appearance of a so-called {\it odd} or Hall viscosity.
%
In general the viscosity, $\eta_{ijkl}$, relates stress, $\sigma_{ij}$, to deformation rate, $\nabla_l v_k$.
When TRS holds, Onsager reciprocal relations (ORR)~\cite{Onsager,NJP,Mazur} for equilibrium fluids require that the standard dissipative viscosity tensor $\eta_{ijkl}^{\rm{e}}$ be even under time reversal (TR) and under the interchange $ij \leftrightarrow kl$.
However, when TRS is broken, ORR predict an additional `odd viscosity' (OV) that is odd under both TR and interchange 
of $ij$ and $kl$: $\eta^o_{ijkl} = -\eta^o_{klij}$. 
This {\it odd viscosity} is non-dissipative and does not contribute to the entropy or heat production.
Hence, it should be present even in a purely non-dissipative Hamiltonian system.  
\begin{figure}
	\begin{centering}
		\includegraphics[scale=0.25]{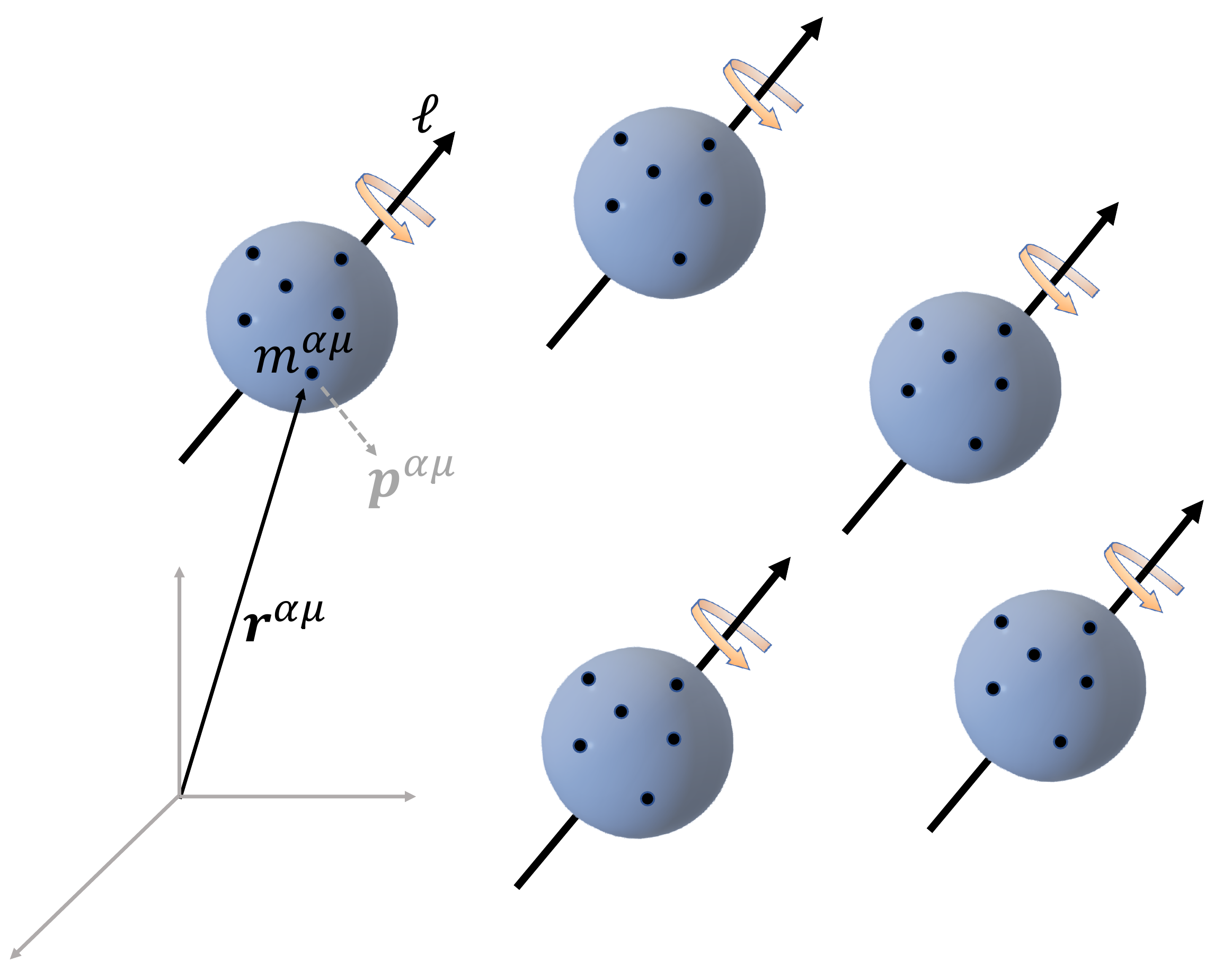}
		\par\end{centering}
	\caption{ Schematic of our model system. Each molecule with CM at $\vecr^\alpha$ (large blue spheres) 
		is composed of many point-like particles (small black spheres) located at $\vecr^{\alpha\mu}$ with mass $m^{\alpha\mu}$
		and momentum $\vecp^{\alpha\mu}$. Each molecule has the same angular momentum ${\bm \ell}^\alpha$, which could be a consequence of {\it e.g.} an external field. The angular momentum breaks rotational symmetry.
	}
	\label{fig:schematic} 
\end{figure}
%

Odd viscosity, often called gyro viscosity, has been studied for some time in gases~\cite{Kagan-Mak} and plasmas in a magnetic field \cite{Braginskii,LLfluids}, 
and in superfluid He$^3$~\cite{He}. It was first discussed by Avron and co-workers in the context of quantum-Hall fluids~\cite{Avron1995} and hypothetical 2D odd fluids~\cite{Avron1998} in which OV is compatible with rotational isotropy.
Subsequently, the effects of OV in quantum-Hall fluids were thoroughly investigated~\cite{Hoyos2014,Read2009,Read2011,Bradlyn2012,Gromov2017}. 
Recent research has paid much less attention to 3D systems.

The fact that TRS is inherently broken in active matter inspired recent investigations 
of OV in classical active materials~\cite{Banerjee2017,Ganeshan2017,Abanov2018,Lapa2014,Lucas2014,Tsai2005}, all of which focused on 2D fluids. 
Most of these studied the phenomenology of OV, though Ref.~\cite{Banerjee2017} derives the OV from an assumed hydrodynamic action
and Ref.~\cite{VitelliKubo} presents a semi-microscopic theory for 2D active chiral fluids in which OV is found numerically using Langevin dynamics simulations. Odd viscosity has also been experimentally confirmed in 2D~\cite{Irvine2019}, verifying the existence of unidirectional edge-states~\cite{Abanov2018,Souslov2019}. 

In this paper we present a first-principles microscopic Hamiltonian theory for odd viscosity in both 2D and 3D using the Poisson-Bracket (PB)
approach~\cite{HohenbergHal1977,Lubensky2003,LubenskyBook}.
Our theory, valid even at zero temperature, suggests that any system with globally or locally aligned spinning components has global (or local) OV arising from kinetic energy alone (other contributions are possible).
We discuss some consequences of OV in 3D fluids, including the breakdown of Bernoulli's principle and 
the existence of anisotropically propagating bulk transverse velocity waves.
We further show that OV should be present in many active matter systems, including 3D ones. 
Specifically, we expect to find signatures of OV in swimming bacteria and actomyosin networks. 
  
We view a fluid as a large collection of particles (rigid molecules), with center-of-mass (CM) position $\vecr^\alpha$, 
each composed of multiple sub-particles (atoms) with mass $m^{\alpha\mu}$ and momentum $\vecp^{\alpha\mu}$
located at $\vecr^{\alpha\mu}$, Fig.~\ref{fig:schematic}.
The total momentum density is $\hat{\bf g}(\vecr) = \sum_{\alpha\mu} {\bf p}^{\alpha\mu}\delta\left(\vecr-\vecr^{\alpha\mu}\right)$,
which, when coarse-grained, becomes~\cite{Supp,Martin}
\begin{eqnarray}
\label{e1}
\vecg(\vecr)  = \vecg^c + \frac{1}{2} {\bm \nabla} \times \vecl  \, , 
\end{eqnarray}
where ${\bf g}^c=\rho \vecv^c$ is the CM momentum density, $\rho$ the mass density, $\vecv^c$ the CM velocity, and $\vecl(\vecr)={\underline{\underline{\boldsymbol{I}}}} \cdot {\bm \Omega}$ is the angular-momentum density of spinning particles, where ${\underline{\underline{\boldsymbol{I}}}}$ is the moment-of-inertia density and ${\bm \Omega}$ is the rotation-rate vector. 
Because by definition $\vecl = 0$ for point-like particles, we necessarily consider complex particles.
Although $\vecl$ affects local momentum both in the bulk and on the surface, 
it does not contribute to the total momentum because $\int\D\vecr\, {\bm \nabla} \times \vecl = 0$.  
It does, however, contribute to the total angular momentum, ${\bm L} = \int \D\vecr \left(\vecr \times {\bf g}^c + \vecl\right)$. 
Like a magnetic field, $\ellv$ is a pseudovector that is even under parity (P) and odd under TR.

Balance of angular momentum implies that the stress tensor ${\bm \sigma}$ associated with $\vecg$ can always be symmetrized~\cite{Martin}. Then, the viscosity $\eta_{ijkl}$ is invariant under the exchanges $i \leftrightarrow j$ and $k \leftrightarrow l$. On the other hand, $\vecg^c$ does not count all momentum, and its stress tensor ${\bm \sigma}^c$  can have antisymmetric contributions. As shown below, a `proper' OV (obeying ORR) appears in ${\bm \sigma}$, while ${\bm \sigma}^c$ contains only one part of it.

In writing~(\ref{e1}) we assume the system is structurally isotropic. However, the presence of $\vecl$ breaks rotational invariance. The essential features of OV are seen for a purely kinetic Hamiltonian, which is written within our model after coarse graining~\cite{Supp,Future} as~\footnote{The complete coarse-grained Hamiltonian includes a term $F[\rho]$, which give rise to an isotropic pressure but do not affect the OV (see Sec. II.A.3 in~\cite{Supp} and~\cite{Lubensky2003,LubenskyBook}}.
\begin{equation}
\label{e2}
H  = \int \D \vecr \, \frac{ \vecg^2 }{2\rho}  = \int \D \vecr\left[  \frac{ (\vecg^c)^2 }{2\rho}  + \ellv\cdot\omegav^c\right]   \, ,
\end{equation}
where ${\bm \omega^c} = \half{\bm \nabla} \times {\bf v}^c$ is half the fluid vorticity and ${\bf g} = \rho{\bf v}$ with $\vecv$ the fluid velocity.
Note that in the second equality we dropped a non-hydrodynamic term $\sim({\bm \nabla} \times \vecl)^2$.
Although this Hamiltonian is standard in terms of $\vecg$, it is peculiar in terms of $\vecg^c$, 
where the second term couples angular momentum density and vorticity. 
This term was added {\it ad-hoc} (with opposite sign) in~\cite{Banerjee2017,Lingam} to a hydrodynamic action and was crucial in deriving the OV.

As detailed in~\cite{Supp}, using the PBs~\cite{Lubensky2005}
\begin{eqnarray}
\label{e3}
\nonumber\{g^c_i(\vecr),\rho(\vecr')\} &=& \rho(\vecr)  \nabla_i \delta(\vecr'-\vecr) \, , \\
\nonumber\{g_i^c(\vecr),g_j^c(\vecr')\} &=& g_i^c(\vecr') \nabla_j \delta(\vecr-\vecr') - \nabla'_i \!\left[ g_j^c(\vecr') \delta(\vecr-\vecr')  \right] , \\
\nonumber\{\ell_i(\vecr),\ell_j(\vecr')\} &=& - \varepsilon_{ijk} \ell_k(\vecr) \delta(\vecr-\vecr') \, ,\\
\{\ell_i(\vecr),g_j^c(\vecr')\} &=& \ell_i(\vecr') \nabla_j \delta(\vecr-\vecr') \, ,
\end{eqnarray}
give the non-dissipative part of the dynamics for the total momentum that satisfies $\dot{g}_i +  \nabla_j \left(v_j g_i\right) = \nabla_j \sigma_{ij}$, where 
\begin{eqnarray}
\label{e4a}
\sigma_{ij} = -P\delta_{ij} + \left( \eta^{e}_{ijkl}  +  \eta^{o}_{ijkl} \right)\nabla_l v_k  \,  ,
\end{eqnarray}
is the complete stress tensor with $P$ the hydrostatic (thermodynamic) pressure. When anisotropic dissipative terms associated with $\vecl$ are ignored, $\eta^{e}_{ijkl} = \lambda \delta_{ij} \delta_{kl}  + \eta \left( \delta_{ik}\delta_{jl} +\delta_{jk}\delta_{il}  \right)$, where $\lambda$ and $\eta$ are constants~\cite{Supp}. The odd viscosity, 
\begin{eqnarray}
\label{e4aa}
& &\eta^{o}_{ijkl} =   -\frac{1}{4} \ell_n \left(  \varepsilon_{jln}\delta_{ik} + \varepsilon_{iln}\delta_{jk} 
+  \varepsilon_{ikn}\delta_{jl} +  \varepsilon_{jkn}\delta_{il} \right) \, , \quad
\end{eqnarray}
naturally emerges from Eqs.~(\ref{e2})-(\ref{e3}), revealing its non-dissipative nature. $\vecl$ can be a function of space and time to create ``localized'' OV~\footnote{Symmetry allows for additional `odd' terms of order ${\cal O}\left(|\vecl|^3\right)$, see~\cite{Supp} Eq.(47).}. When $\vecl$ is constant we get Onsager's OV, where $\vecl$ plays the role of magnetic fields in plasmas~\cite{Braginskii,LLfluids}. The 2D OV~\cite{Banerjee2017,Avron1998} follows from Eq.~(\ref{e4aa}) by writing $\vecl = \ell \hat{z}$ (for a fluid in the $xy$ plane), thereby converting the Levi-Civita tensor to $\varepsilon_{ij} = \varepsilon_{ijz}$. Remarkably, unlike magnetic field in plasmas, this result is purely mechanical and does not require thermodynamic equilibrium or statistical mechanics.

We continue with some phenomenological consequences of OV in 3D. We focus, for simplicity, on the case of constant (in space and time) $\vecl$, which may originate in external  driving as described in~\cite{Supp} and Refs.~\cite{Tsai2005,Banerjee2017,Bartolo2016}), and experimentally realized in 2D ``fluid'' metamaterials~\cite{Tsai2005,Irvine2019}. The `odd' Navier-Stokes equation (NSE), Eq.~(\ref{e4a}), then reads
\begin{eqnarray}
\label{e6}
\nonumber\dot{g}_i +  \nabla_j \left(v_j g_i\right) &=& -\nabla_i \tilde P + \eta \nabla^2 v_i 
 +\frac{1}{2} {\bm \ell}\cdot \nabla \omega_i  \\
&-& \half\varepsilon_{ikn} \ell_n \nabla_k \nabla \cdot \vecv  \, ,
\end{eqnarray}
with $\tilde{P} \equiv  P  - (\lambda + \eta) \nabla\cdot\vecv + \vecl \cdot {\bm \omega}$ being the mechanical pressure (diagonal part of the stress) and $\omegav=\half{\bm \nabla} \times \vecv$.
To investigate the mode structure we linearize Eq.~(\ref{e6}) and the continuity equation, $\dot{\rho} + \nabla \cdot (\rho\vecv) = 0$, and 
use Fourier-Transform $\left[{\bf v},\Drho\right] = \int \frac{\D \bf k}{(2\pi)^3}\left[\tilde{\bf v},\delta\tilde\rho\right] \e^{i\left({\bf k}\cdot\vecr - s t\right)}$ to obtain:
\begin{eqnarray}
\label{ms1a}
\begin{pmatrix}
s + i\nu k^2  					  & -i\tilde\ell kk_\parallel   & 2i\tilde\ell kk_\perp  	&  0\\
i\tilde\ell k k_\parallel       & s + i\nu k^2 \,\,  	        &	0		        				& 0 \\
-2 i\tilde\ell kk_\perp        &	0 						    	& s + i\nu_L k^2       		 &  -kc \\
0 								   	   &	0 					    		&		  -kc 				 	   &  s
\end{pmatrix} 
\begin{pmatrix}
\tilde{v}_1	\\
\tilde{v}_2 \\
\tilde{v}_L \\
\tilde{h}
\end{pmatrix} = 0 \, . \quad  
\label{eq:dynamics}
\end{eqnarray}
Here $\nu \equiv \eta/\rho_0$, $\nu_L \equiv 2\nu + \lambda/\rho_0$,  $\tilde\ell \equiv \ell/(4\rho_0)$, $\vecl = \ell \hat{z}$, 
and $h = \Drho/(c \rho_0)$ where $\rho = \rho_0 + \Drho$ and $c$ is the sound speed ($c^2=\partial P/\partial \rho$).
We further define $v_{1,2} \equiv \vecv \cdot \hat{\bf e}_{1,2}$ and $v_L \equiv \vecv \cdot \hat{\bf k}$,
where $\hat{\bf k} = (k_x,k_y,k_z)/k$, $\hat{\bf e}_1 = (-k_y,k_x,0)/k_\perp$, 
and $\hat{\bf e}_2 = (-k_xk_z,-k_yk_z,k_\perp^2)/(kk_\perp)$ are a set of orthonormal vectors with
$k = \sqrt{{\bf k}^2}$, $k_\perp = \sqrt{k_x^2 + k_y^2}$, and $k_\parallel = {\bf k} \cdot \vecl / \ell$.

\begin{figure}
	\begin{centering}
		\includegraphics[width=0.4\textwidth]{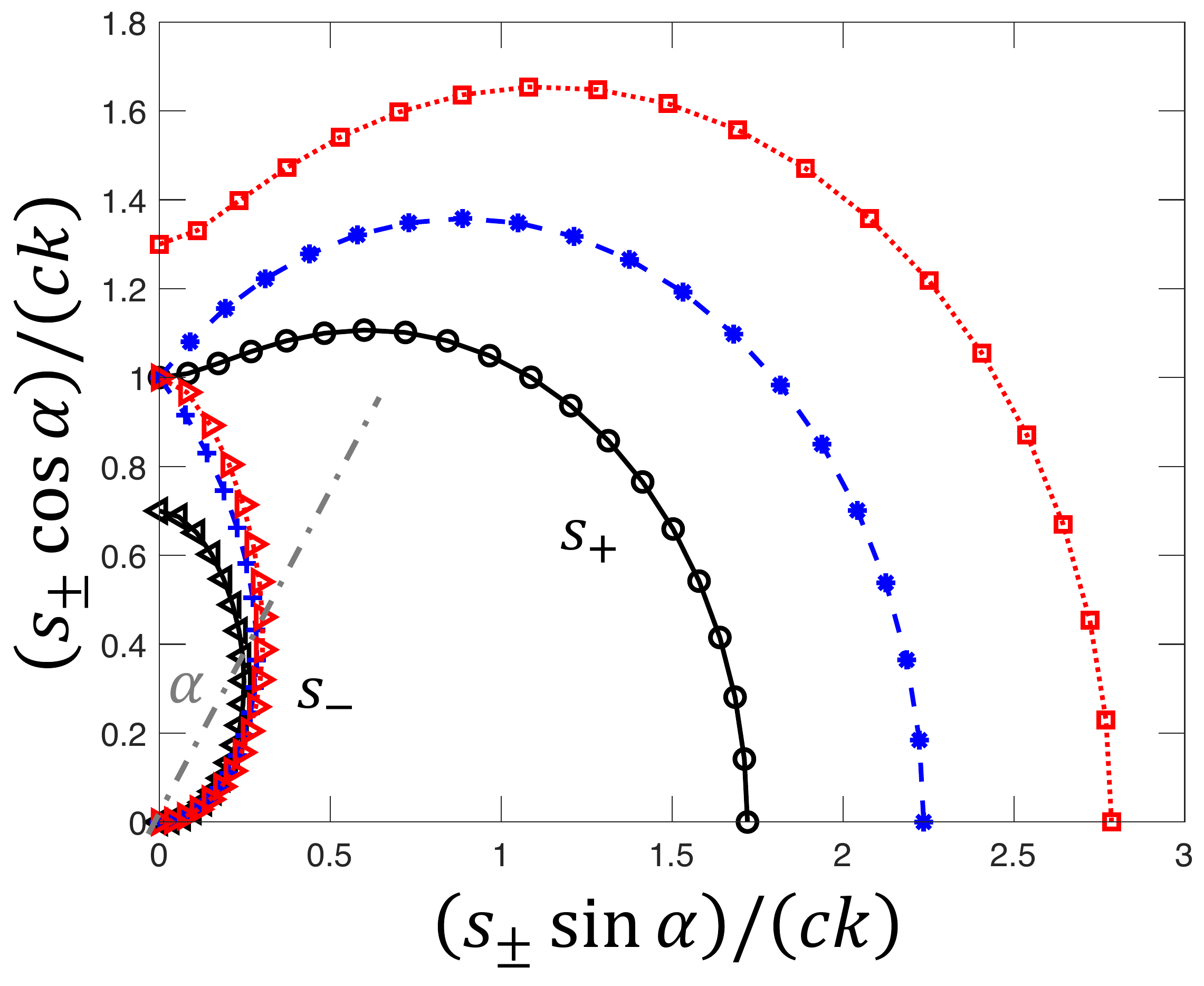}
	\end{centering}
	\caption{
		A polar plot of the two normalized `odd' excitations as function of the angle $(\alpha)$ between $\vecl$ and ${\bf k}$.
		Black lines (reverse triangle for $s_-$ and circles for $s_+$) are for $k\tilde\ell/c = 0.7$,
		blue lines (plus for $s_-$ and asterisk for $s_+$) are for $k\tilde\ell/c  = 1$,
		and the red lines (triangles for $s_-$ and squares for $s_+$) are for $k\tilde\ell/c  = 1.3$.
		These anisotropic excitations are similar to those found in columnar liquid crystals~\cite{LC1980} 
		and nematic elastomers~\cite{Stenull2004}.
	}  
	
	\label{fig:sound}
\end{figure}

When dissipation ($\nu$, $\nu_L$) is negligible, the matrix in Eq.~(\ref{ms1a}) is Hermitian, 
with real  eigenvalues, and corresponding orthogonal eigenvectors.
Hence, there are no instabilities and novel `odd' mechanical waves always propagate. The full solution for the inviscid case can be found in~\cite{Supp} (Eq.~(81)).
Figure~\ref{fig:sound} presents a polar plot of the `odd' excitation frequencies $s_{\pm}$ for various $\tilde{\ell}$.
We find that $s_-$ ($s_+$) reaches its maximal value for $\alpha = 0$ ($\alpha = 45^\circ$), where $\vecl\cdot{\bf k} = \cos\alpha$.
These modes are always admixture of transverse (T) and longitudinal (L) modes, except when $\alpha=0$, in which case, $s_+$ ($s_-$) is the L mode for $k\tilde\ell<c$  ($k\tilde\ell>c$).   

The upper left (lower right) $2\times 2$ submatrix of Eq.~(\ref{eq:dynamics}) deals with transverse (longitudinal) variables only, where $2\tell k k_{\perp}$ couples the two.  The dimensionless measure of this quantity is $g= 2 \tell k_{\perp}/c$, which becomes arbitrarily small as $k_\perp$ becomes less than $c/(2 \tell)$. Thus, in the hydrodynamic limit, T and L modes decouple. The frequencies with lowest-order coupling corrections are
\begin{align}
s_T & = (\pm \tell k k_\parallel - i \nu k^2)(1- 2 \tell^2 k_{\perp}^2/c^2) \label{eq:sT}\\
s_L & = \pm c k \left(1 + 2 \frac{\tell^2 k_{\perp}^2}{c^2} - \frac{1}{8}\frac{\nu_L k^2}{c^2} \right)- \frac{1}{2} i \nu_L k^2 \, ,
\label{eq:sL}
\end{align}
and the associated eigenvectors [$\vec{v}=(v_1,v_2,v_L,h)$] to lowest order in $g$ and $g_\nu = k\nu_L / c$ are:
$\vec{v}_L  = (\mp i g, 0,1,\pm1 + ig_\nu/2)$ and $\vec{v}_T  = (1, \mp i, 0, - i g )$.
As expected~\cite{Avron1998}, the transverse polarizations are purely circular in the $g \rightarrow 0 $ limit. When dissipative viscosities are nonzero, the transverse-mode frequency becomes purely diffusive for $k_\parallel=0$. 
A detailed analysis of the general case with arbitrary $\bm k$ and the associated phase diagram showing regions 
with diffusing and propagating modes is deferred to~\cite{Supp}.
%

It is instructive to see how breaking TRS and parity by $\vecl$ affects normal modes and field couplings.  
Define the signature of a field to be $({\rm TR},{\rm P})$.
$v_1$ and $v_L$ have signature $(-,+)$, $v_2$ $(-,-)$, and $\tell$  $(-,+)$. 
Thus, $\tell k k_\parallel$ with signature $(-,-)$ couples $\dot{v}_1$ to $v_2$, 
creating a propagating transverse mode, and $\tell k k_\perp$ with signature $(-,+)$ couples $\dot{v}_1$ to $v_L$, thus mixing longitudinal and transverse components. 
%
The transverse $v_1-v_2$ block in Eq.~(\ref{eq:dynamics}) is similar to the equation for magnons in ferromagnets~\cite{Kittel} with magnetization  ${\mathbf{M}=(M_x,M_y,M_z)}$ 
with signature $(-,+)$. $M_x$ and $M_y$ play the role of $v_1$ and $v_2$, and $M_z$ the role of $\tell$. 
The magnon has an isotropic, rather than an anisotropic, dispersion relation $\omega \propto k^2$, reflecting the fact that in the absence of spin-orbit couplings 
$\mathbf{M}$ rotates under a different group than spatial points.

%
%

An important and useful simplification of the odd NSE (Eq.~(\ref{e6})) is its incompressible limit, ${\bm \nabla} \cdot {\bf v} = 0$:
\begin{eqnarray}
\label{e7}
\dot{g}_i +  \nabla_j \left(v_j g_i\right) = -\nabla_i  \tilde{ P}  + \eta \nabla^2 v_i +  \frac{1}{2} {\bm \ell}\cdot \nabla \omega_i    \, ,
\end{eqnarray}
where now $\tilde{P} \equiv  P  + \vecl \cdot {\bm \omega} $.
Notice that in 2D the last term vanishes, and because $\tilde{P}$ is not a state variable but rather is obtained from the incompressibility constraint,
the (bulk) flow is not affected by OV~\cite{Ganeshan2017}.
However, this is not the case in 3D, thus one should expect very different phenomenology from that of 2D, even for incompressible fluid.

In order to understand the incompressible limit it is useful to examine the longitudinal equation.
The latter is obtained by taking the divergence Eq.~(\ref{e6}) and using the continuity equation:
%
\begin{eqnarray}
\label{e10}
\left[ \partial_t^2   - \frac{ \lambda+2\eta}{\rho_0}\nabla^2 \partial_t  \right]  \Drho -\nabla^2 \left( P + {\bm \ell} \cdot {\bm \omega}  \right)   =0  \, .
\label{eq:incomp}
\end{eqnarray}
In normal fluids ($\vecl = 0$), incompressibility is attained by setting $\Drho = 0$, implying that $\nabla^2 P = 0$. 
This constraint is not affected by transverse diffusion modes or by local vorticity. 
In the present case, the constraint $\Drho =0$ implies that $\nabla^2(P + {\bm \ell} \cdot {\bm \omega})=0$. 
But in the presence of a transverse wave, in which ${\bm \ell} \cdot {\bm \omega}$ is nonzero, $P$ must undergo a compensating change, 
which normally means that $\Drho$ must do so as well.
The resolution requires consideration of the limit $c \rightarrow \infty$.  
The eigenvectors $\vec{v}_T$ and $\vec{v}_L$ following Eq.~(\ref{eq:sL}) reveal that $\Drho = -  {\bm \ell} \cdot {\bm \omega}/c^2$,
which satisfies Eq.~(\ref{e10}) for $c\to\infty$~\cite{Supp}.
Because $\Drho$ is formally zero, but $\nabla^2 P$ is not, we could say that a transverse wave produces a (thermodynamic) pressure wave, but not a density wave.
Note that although $P$  oscillates, the experimentally accessible $\tilde P$ does not.
This shows the crucial difference between these two definitions of pressure.

We continue by examining the validity of Bernoulli's principle in odd fluids. 
Consider an incompressible, inviscid ($\eta = 0$) odd fluid. 
In steady-state, multiplying Eq.~(\ref{e7}) by ${\bf v}$ gives
%
\begin{eqnarray}
\label{e8}
\left({\bf v} \cdot \nabla\right) \left( \frac{\tilde{P}}{\rho} +\half v^2 \right) = \frac{1}{2\rho} v_i  \left( \vecl \cdot {\bm \nabla} \right) \omega_i    \, .
\end{eqnarray}
As in 2D, the mechanical pressure $\tilde{P}$ takes the place of the thermodynamic pressure $P$.  
In 2D, the right-hand-side (RHS)  of Eq.~(\ref{e8}) vanishes and Bernoulli's principle, in which $\tilde{P} + \half v^2$ is constant along a streamline, is recovered.
Furthermore, in 2D the vorticity is conserved along a streamline~\cite{acheson}, and thus Bernoulli's principle for $P$ is also restored~\cite{Avron1998}.
In 3D the RHS does not generally vanishes, therefore, Bernoulli's principle is not valid for `odd' fluids in 3D.
Similarly, we observe the breakdown of Kelvin's circulation theory in 3D odd fluids~\cite{Supp}, and expect to find significant effects on lift and the Magnus effect.


So far we have not discussed the origin of a non-vanishing angular momentum, which requires discussing the dynamics of $\vecl$.
The PB approach along with introduction of a torque density ${\bm \tau}(\vecr,t)$~\footnote{$\tauv$ can include a rotational friction term $-\Gamma^\Omega\Omegav$, see~\cite{Supp} Sec. IV.A.}, provides the dynamics' non-dissipative part~\cite{Supp}. Adding a dissipative term $-\Gamma_{ij}(\Omega_j-\omega_j^c)$~\cite{NJP,Lubensky2005} that provides preference for a dissipation-free steady-state in which ${\bm \Omega} = {\bm \omega^c}$ completes the equation:
\begin{equation}
\label{e5}
\dot{\ell}_i(\vecr) + \nabla_j\left(\ell_i v_j\right) = \varepsilon_{ijk}\omega_j \ell_k- \Gamma \left(\Omega_i - \omega_i\right) + \tau_i  \, ,
\end{equation}
where a non-hydrodynamic term $\sim\nabla^2 \vecl$ was neglected, and we set for simplicity $\Gamma_{ij} = \Gamma \delta_{ij}$. The 2D version of Eq.~(\ref{e5}), in which the first term of the RHS vanishes, was first suggested in~\cite{Tsai2005} and was used extensively since~\cite{Bartolo2016,NJP,Banerjee2017,Mazur}.
Both TRS and parity are broken by $\vecl$, and the presence of ${\bm \tau}$ adds an extra term $\half\varepsilon_{ijk} \tau_k$ to $\sigma_{ij}$~\footnote{If $\tauv$ is constant in space, this term does not affect the excitation spectrum of Eqs.~(\ref{eq:sT})-(\ref{eq:sL}).}, giving $\vecl$ a non-random value, $\vecl \simeq \underline{\underline{\bf I}} \cdot {\bm \tau} / \Gamma$ in the hydrodynamic limit (see~\cite{Supp} Eq.~(42)) leading to the appearance of OV, Eq.~\eqref{e4aa}. When ${\bm \tau } = 0$, $\Omegav$ relaxes in microscopic times to a function of $\bm\omega$, which rapidly vanishes as the fluid approaches equilibrium.

Equation~(\ref{e4a}) gives the dynamics of $\vecg$, for which a symmetric stress tensor can always be found (up to $\half\varepsilon_{ijk} \tau_k$ that appears in the presence of body torques). The dynamics of $\vecg^c$ assumes a similar form~\cite{Supp}, $\dot{g}^c_i +  \nabla_j \left(v^c_j g^c_i\right) = \nabla_j \sigma^c_{ij}$,  but with
\begin{eqnarray}
\label{e5a}
\nonumber\sigma^c_{ij} &=& -P\delta_{ij} + \left[ \eta^e_{ijkl} - \half\ell_n\varepsilon_{jkn}\delta_{il} \right] \nabla_l v^c_k \\
&-& \half v^c_i \left(\nabla\times\vecl\right)_j+ \frac{\Gamma}{2} \varepsilon_{ijk} \left(\Omega_k - \omega^c_k\right) \, ,
\end{eqnarray}
that has an antisymmetric dissipative part to ensure conservation of angular momentum when $\tauv=0$, and both symmetric and antisymmetric reactive parts, but only one part of Onsager's OV (Eq.~\eqref{e4aa}).  This contrasts with current literature~\cite{Banerjee2017,Souslov2019,VitelliKubo} in which the CM stress tensor ${\bm \sigma}^c$ is assumed to include both the `proper' OV and the antisymmetric dissipative term~\footnote{Other contributions to the OV ({\it e.g.} arising from interactions) may produce additional `proper' OV in ${\bm \sigma}^c$, but it cannot cancel the non-interacting contribution discussed here.}.

In~\cite{Supp} we show that after integrating out ${\bm \Omega}$, the CM stress tensor has the proper Onsager OV, terms proportional to $\nabla \times \vecl$, and an additional reactive antisymmetric viscosity contribution $\half \ell_k \varepsilon_{ijk} \nabla\cdot\vecv$. The latter violates ORR and implies that $\dot\vecl\neq 0$ (see~\cite{Supp}). The corollary is that even after relaxation of ${\bm \Omega}$, the CM description cannot properly describe a system with $\vecl\neq 0$, because it does not conform with the balance of angular momentum (an exception is a fluid of constant density).

\begin{figure}
	\begin{centering}
		\includegraphics[scale=0.25]{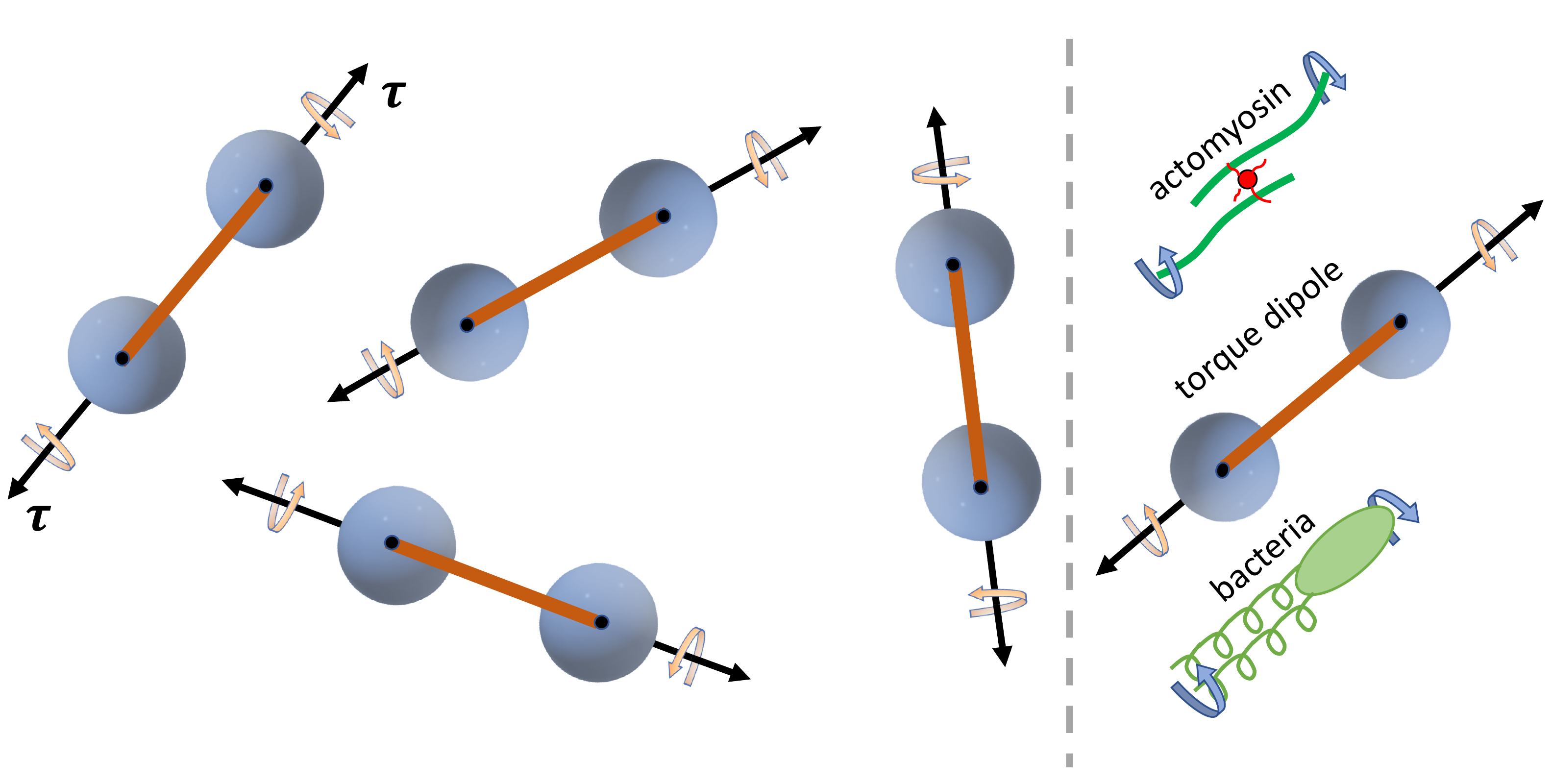}
		\par\end{centering}
	\caption{ Illustration of a fluid of torque dipoles that shows inhomogeneous OV.
		On the right - torque dipole as a model for toque exerted by bacteria (where its head and flagellar are rotating in opposite directions)
		and for a myosin twisting two actin filaments~\cite{NJP}.		}
	\label{fig:schematic2} 
\end{figure}

Although we explore in detail the constant $\vecl$ case, our framework does not restrict $\ellv$ to be constant in space or time.
In fact, in internally driven active materials (such as all natural active materials) the total active torque must vanish~\cite{NJP}, 
implying that ${\bm \tau}$ is a divergence of some quantity. For example, in an isotropic active gel ({\it e.g.} actomyosin gel) ${\bm \tau} = \tilde\tau {\bm \nabla} \rho$ 
where $\tilde\tau$ is a pseudoscalar (see derivation in~\cite{Supp} Sec.~V).
Therefore, we expect to have numerous realizations of 3D OV 
ranging from swimming bacteria~\cite{Elsen2017} and actomyosin networks~\cite{NJP} to swimming microrobots (see Fig.~\ref{fig:schematic2}).

We have presented a microscopic Hamiltonian theory for the appearance of {\it odd viscosity} in active fluids. 
Being a Hamiltonian theory, no dissipation is required to obtain the OV terms. Our central result is an equation of motion for the total momentum density of non-interacting spinning particles that is valid for arbitrary local values of the angular momentum density $\vecl$.  This equation, which is the analogue of Bloch equations for magnetization, yields the OV predicted by Onsager. Interactions among particles, which we do not consider, modify the OV value but not its form.

Our work considerably extends  the applicability of OV into 3D systems and specifically shows its relevance in biological realizations and systems in which torques are generated internally. Examples for such biological realizations range from bacterial suspensions to actomyosin networks, and may even be present in active biopolymer networks without motors~\cite{sihan} where filaments chirality couples force and twist. 
It is our hope this work will promote a variety of studies on odd viscosity in 3D systems, from ferronematics to active gels.
For instance, it would be interesting to investigate the appearance of odd viscosity in the presence of nematic or polar order.

\begin{acknowledgments} 
\emph{Acknowledgments:} 
TM thanks Fred MacKintosh for fruitful discussion.
TM acknowledges funding from the National Science Foundation Center for Theoretical Biological Physics (Grant PHY-2019745) and TCL acknowledges funding from the National Science Foundation Materials Research Science and Engineering Center (MRSEC) at University of Pennsylvania (grant no. DMR-1720530). TCL's work on this research was supported in part by the Isaac Newton Institute
for Mathematical Sciences during the program ``Soft Matter Materials - Mathematical Design Innovations''
and by the International Centre for Theoretical Sciences (ICTS) during the program - ``Bangalore School
on Statistical Physics - X (Code: ICTS/bssp2019/06).''
\end{acknowledgments}

\end{document}


\title{Supplementary Material \\  Odd viscosity in active matter: microscopic origin and 3D effects}

\author{Tomer Markovich$^{1}$}
\email{tm36@rice.edu} 
\author{Tom C. Lubensky$^{2}$}
\affiliation{
	$^{1}$Center for Theoretical Biological Physics, Rice University, Houston, TX 77005, USA \\
	$^{2}$Department of Physics and Astronomy, University of Pennsylvania, Philadelphia, Pennsylvania 19104, USA
}

\date{\today}

\maketitle

\section{Microscopic model for the total momentum}

A main result cited in Eq.~(1) of the main text relates the total momentum density to the angular momentum density. 
Here we derive this equation using a very general microscopic model and a simplistic diatomic molecule model.
The crucial ingredient to produce Eq.~(1), of any microscopic model, is the complexity of the individual molecules, 
{\it i.e.} they cannot be simply point masses, but must contain at least `mass dipole' (diatomic molecule).

\subsection{General model}

Let us start with a general model describing complex molecules.
We consider a fluid that is composed of of many complex rigid particles~\footnote{Rigidity is assumed for simplicity - internal degrees of freedom will not change our hydrodynamic equations.} with center-of-mass (CM) at position $\vecr^\alpha$, 
each of which is by itself comprised of two or more point-like mass points with mass $m^{\alpha\mu}$ and momentum 
$\vecp^{\alpha\mu} = m^{\alpha\mu} \dot\vecr^{\alpha\mu}$, located at $\vecr^{\alpha\mu}$ (see Fig.~1 of main text).
Throughout we use $\dot X = \partial x / \partial t$.
The total momentum density of such a fluid is:
\begin{eqnarray}
\label{m1}
\hat{g}_i(\vecr) =\sum_{\alpha\mu} p_i^{\alpha\mu} \delta \left(\vecr - \vecr^{\alpha\mu}\right) 
\simeq \sum_\alpha p_i^\alpha \delta(\vecr-\vecr^\alpha) - \sum_{\alpha\mu} p_i^{\alpha\mu} \Delta r_j^{\alpha\mu}  \nabla_j \delta(\vecr-\vecr^\alpha)\, ,
\end{eqnarray}
where $\Delta\vecr^{\alpha\mu} \equiv \vecr^{\alpha\mu} - \vecr^\alpha$ and we define the CM momentum $\vecp^\alpha \equiv \sum_\mu \vecp^{\alpha\mu}$.
Here we assume the molecules are small compare to distances of interest (long wavelength limit), $|\Delta\vecr^{\alpha\mu}| \ll |\vecr-\vecr^\alpha|$.
The second term in Eq.~(\ref{m1}) can be written in a more familiar form with,
\begin{eqnarray}
\label{m2}
\nonumber\sum_{\mu} p_i^{\alpha\mu} \Delta r_j^{\alpha\mu} &=& \sum_{\mu} m^{\alpha\mu} \Delta r_j^{\alpha\mu} \partial_t \Delta r_i^{\alpha\mu} 
+ \sum_{\mu}  m^{\alpha\mu} \dot{r}^\alpha_i \Delta r_j^{\alpha\mu} 
= \sum_{\mu} m^{\alpha\mu} \Delta r_j^{\alpha\mu} \partial_t \Delta r_i^{\alpha\mu} 
+ \dot{r}^\alpha_i  \left[  \sum_\mu m^{\alpha\mu} r_j^{\alpha\mu} - r_j^\alpha\sum_\mu m^{\alpha\mu} \right] \\
&=& \sum_{\mu} m^{\alpha\mu} \Delta r_j^{\alpha\mu} \partial_t \Delta r_i^{\alpha\mu}  \, ,
\end{eqnarray}
where we have used the fact that $\vecr^\alpha$ is the CM coordinate, {\it i.e.}, $\vecr^\alpha \sum_\mu m^{\alpha\mu} = \sum_\mu m^{\alpha\mu} \vecr^{\alpha\mu}$.

The `spin' momentum density is defined to be the difference between the total momentum density and the CM momentum density, $\vecg^c(\vecr)$:
\begin{eqnarray}
\label{m3}
& & \hat{g}^s_i(\vecr) \equiv \hat{g}_i(\vecr) - \hat{g}_i^c(\vecr) = -\sum_{\alpha\mu} m^{\alpha\mu}  \left( \partial_t \Delta r_i^{\alpha\mu} \right) \Delta r_j^{\alpha\mu}
\nabla_j \delta(\vecr-\vecr^\alpha) \, , \\
& &\hat{\vecg}^c(\vecr) \equiv \sum_\alpha \vecp^\alpha \delta(\vecr-\vecr^\alpha) \, .
\end{eqnarray}
%
%
Separating $\vecg^s$ into symmetric and antisymmetric parts (for $i \leftrightarrow j$) we have,
\begin{eqnarray}
\label{m4}
\hat{g}_i^s(\vecr) = -\frac{1}{2} \Bigg[ \varepsilon_{ijk} \varepsilon_{klm} \sum_{\alpha\mu} m^{\alpha\mu} \left( \partial_t \Delta r_l^{\alpha\mu} \right) \Delta r_m^{\alpha\mu} 
 \nabla_j \delta(\vecr-\vecr^\alpha) 
+ \frac{3}{2} \sum_\alpha \left( \dot{R}_{ij}^\alpha +\frac{1}{3}\dot{I}_{ij}^\alpha \right)  \nabla_j\delta(\vecr-\vecr^\alpha)   \Bigg] \, ,
\end{eqnarray}
where we use the definitions of ${\bf R}^\alpha$ and ${\bf I}^\alpha$ from Ref.~\cite{Lubensky2003}:
\begin{eqnarray}
\label{m5}
& &R_{ij}^\alpha = \sum_{\mu} m^{\alpha\mu} \left[ \Delta r_i^{\alpha\mu} \Delta r_j^{\alpha\mu} - \frac{1}{3}\left(\Delta \vecr^{\alpha\mu}\right)^2\delta_{ij} \right] \, , \\
& &I_{ij}^\alpha = \sum_{\mu} m^{\alpha\mu} \left[ \left(\Delta \vecr^{\alpha\mu}\right)^2\delta_{ij}  - \Delta r_i^{\alpha\mu} \Delta r_j^{\alpha\mu} \right] \, .
\end{eqnarray}
%
%
The first term in the brackets of Eq.~(\ref{m4}) includes a cross product of the positions of the molecule's point masses and their velocities, both relative to the molecule CM.
This cross product is also multiplied by the mass, providing the definition of the angular momentum of each molecule,
\begin{eqnarray}
\label{m6}
\ell^\alpha_i = \varepsilon_{ijk} \sum_{\mu} m^{\alpha\mu} \left( \partial_t \Delta r_k^{\alpha\mu} \right) \Delta r_j^{\alpha\mu} \, .
\end{eqnarray}
Defining the density of angular momentum, $\hat{\vecl}(\vecr) = \sum_\alpha \vecl^\alpha \delta(\vecr-\vecr^\alpha)$, and its symmetric counterpart
$\hat{{\bf A}}(\vecr) = \half\sum_\alpha \dot{\bf K}^\alpha \delta(\vecr-\vecr^\alpha)$ where ${\bf K}^\alpha = \frac{3}{2} \left( {\bf R}^\alpha +\frac{1}{2}{\bf I}^\alpha \right)$, 
we finally obtain
\begin{eqnarray}
\label{m7}
\hat\vecg^s(\vecr) = \half \nabla \times \hat\vecl  + \nabla \cdot \hat{\bf A}  \, .
\end{eqnarray}
As we show below using a diatomic molecule model, $\hat {\bf K}(\vecr) = \sum_\alpha {\bf K}^\alpha \delta\left(\vecr-\vecr^\alpha\right)$ is a generalization of the alignment tensor and should be thought of as an order parameter.

To continue we coarse-grain all fields~\cite{tobepub} to obtain spatially smooth functions. The coarse-grained fields, $X(\vecr)$, are the spatial and temporal averages of the microscopic fields, $\hat{X}(\vecr)$ (average over particles in some mesoscopic volume and over microscopic timescales).
The system is assumed to be isotropic (other than the possible appearance of angular momentum). 
Therefore, in the derivation of the Poisson-Brackets below and throughout (except for in the derivation of the diatomic model) we set ${\bf A} = {\bf K} = 0$, 
and $\vecg^s(\vecr) = \half \nabla \times \vecl$.
This gives Eq.~(1) of the main text:
\begin{eqnarray}
\label{m7a}
{\bf g}(\vecr)  = {\bf g}^c + \frac{1}{2} {\bm \nabla} \times \vecl  \, .
\end{eqnarray}
Note that in general $\ell_i = I_{ij} \Omega_j = \rho \tilde{I}_{ij} \Omega_j$ where $\tilde{\bf I}$ is the moment of inertia per unit mass. For structurally isotropic system $\vecl = \rho \tilde{I} {\bm \Omega}$. Note that the total angular momentum density has the property that it gives the correct total angular momentum of the system,
\begin{eqnarray}
\label{m8}
\nonumber{\bf L} &=& \int \D\vecr \, \varepsilon_{ijk} r_j g_k = \int\D\vecr\left[ \varepsilon_{ijk}r_j g_k^c + \half \varepsilon_{ijk} \varepsilon_{klm} r_j \nabla_l \ell_m  \right] 
= \int\D\vecr \left[  \varepsilon_{ijk} r_j g_k^c  - \half  \varepsilon_{ijk}  \varepsilon_{kjm} \ell_m \right] \nonumber \\
&=&\int\D\vecr \left[ \left(\vecr \times {\bf g}^c \right)_i + \ell_i \right] \, 
\end{eqnarray}
where the integral is over a volume larger than the system's.

Following similar calculation for the density we find that 
\begin{eqnarray}
\label{m9}
\hat{\rho}(\vecr) = \sum_{\alpha\mu} m^{\alpha\mu} \delta\left(\vecr-\vecr^{\alpha\mu}\right) \simeq \hat{\rho}_c(\vecr) + \nabla_i \nabla_j \hat{K}_{ij}(\vecr) \, ,
\end{eqnarray}
with $\hat\rho^c(\vecr) \equiv \sum_\alpha m^{\alpha} \delta(\vecr-\vecr^\alpha)$ and $m^\alpha = \sum_{\mu\in\alpha} m^\mu$ is the mass of the $\alpha$'s molecule.
Therefore, in general $\rho$ depends on the order parameter ${\bf K}$, but as we noted above we consider only the case of ${\bf K} = 0$ so we have $\rho(\vecr) = \rho^c(\vecr)$.

\subsection{Diatomic molecule as a simple case}

\begin{figure}
	\begin{centering}
		\includegraphics[width=0.3\textwidth]{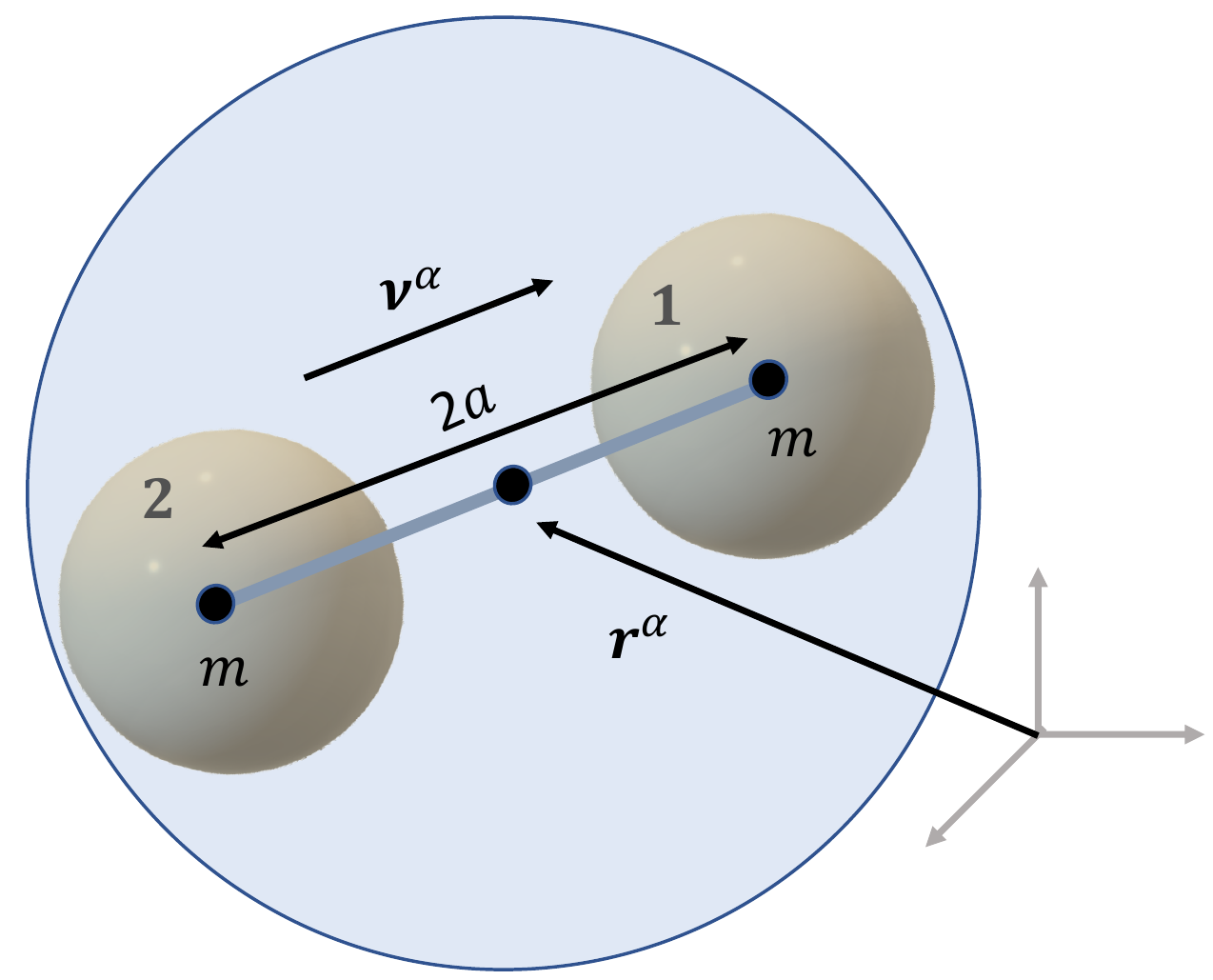}
		\par\end{centering}
	\caption{ Illustration of a diatomic molecule composed of two equal point masses, $m$, separated by distance $sa$.
		The vector ${\bm \nu}$ points from mass `2' to mass `1', and $\vecr^\alpha$ points to the CM of the $\alpha$'s molecule.
	}
	\label{fig:diatomic} 
\end{figure}

To get some more intuition on the microscopic model, we introduce a simple model in which each rigid molecule is composed
of two point masses, $m$, separated by distance $2a$ as depicted in Fig.~\ref{fig:diatomic}.
We define ${\bm \nu}^\alpha$ to be a unit vector pointing from mass `2' to mass `1'.
The momentum of the two point masses can we written in terms of the CM momentum, $\vecp^\alpha$, and ${\bm \nu}$: 
\begin{eqnarray}
\label{e2a}
& &\vecp^{\alpha,1} = \half\vecp^\alpha + ma\dot{\bm\nu}^\alpha  \, , \\
& &\vecp^{\alpha,2} = \half\vecp^\alpha - ma\dot{\bm\nu}^\alpha  \, .
\end{eqnarray}
The total momentum density for the diatomic fluid is
\begin{eqnarray}
\label{e2}
\hat{g}_i(\vecr) &=& \sum_{\alpha\mu} p_i^{\alpha\mu} \delta \left(\vecr - \vecr^{\alpha\mu}\right) 
=\sum_\alpha \left[ p_i^{\alpha,1} \delta(\vecr-\vecr^\alpha - a{\bm \nu}^\alpha) + p_i^{\alpha,2}  \delta(\vecr-\vecr^\alpha + a{\bm \nu}^\alpha)   \right] \\
\nonumber&\simeq& \sum_\alpha \left[\left(p_i^{\alpha,1} + p_i^{\alpha,2}\right)\delta(\vecr-\vecr^\alpha) + a \left(p_i^{\alpha,2} - p_i^{\alpha,1}\right) \nu_j^\alpha \nabla_j  \delta(\vecr-\vecr^\alpha) \right]
=  \sum_\alpha \left[p_i^\alpha \delta(\vecr-\vecr^\alpha) - m\tilde{I}\dot{\nu}_i^\alpha \nu_j^\alpha \nabla_j  \delta(\vecr-\vecr^\alpha)  \right] \, ,
\end{eqnarray}
where we have used $\vecr^{\alpha,1} - \vecr^{\alpha,2} = 2a{\bm \nu}^\alpha$, and the moment of inertia of each molecule is $\tilde{I}=2a^2$.
As before we use the long wavelength limit, $a \ll |\vecr-\vecr^\alpha|$.

Following similar steps as in the general model we write $\hat\vecg = \hat\vecg^c + \hat\vecg^s$, with $\hat\vecg^c \equiv \sum_\alpha \vecp^\alpha \delta(\vecr-\vecr^\alpha)$ and
\begin{eqnarray}
\label{e4}
\hat{g}_i^s(\vecr) = -\frac{m\tilde{I}}{2} \left[ \varepsilon_{ijk} \varepsilon_{klm} \sum_\alpha  \dot{\nu}_l^\alpha\nu_m^\alpha \nabla_j\delta(\vecr-\vecr^\alpha) 
+ \sum_\alpha \dot{Q}_{ij}^\alpha  \nabla_j\delta(\vecr-\vecr^\alpha)   \right] \, .
\end{eqnarray}
Here $Q_{ij}^\alpha = \nu_i^\alpha \nu_j^\alpha - \delta_{ij}/d$ is the alignment tensor and $d$ is the number of dimensions.
As in the general model, the first term of Eq.~(\ref{e4}) includes the angular momentum of each molecule, 
$\vecl^\alpha = m\tilde{I} {\bm \nu}^\alpha \times \dot{\bm \nu}^\alpha$.
The advantage of this simple model is now revealed by using the definition of the angular velocity of each molecule, 
$\Omega^\alpha \equiv {\bm \nu}^\alpha \times \dot{\bm \nu}^\alpha$,
which gives explicitly $\vecl^\alpha = m\tilde{I} {\bm \Omega}^\alpha$.
Calculating the density within this model gives $\rho(\vecr) = \rho^c(\vecr) + m\tilde{I} \nabla_i\nabla_j Q_{ij}(\vecr)$, where $Q_{ij}(\vecr) = \sum_\alpha Q_{ij}^\alpha\delta(\vecr-\vecr^\alpha)$.
Therefore, we conclude that in this simple model ${\bf K}^\alpha \to m\tilde{I} {\bf Q}^\alpha$, and as in the general model we take ${\bf A} = {\bf K} = 0$
such that $\vecg^s(\vecr) = \half \nabla \times \vecl$ and $\rho(\vecr) = \rho^c(\vecr)$.

\section{Poisson Brackets}

The microscopic dynamics obeys Hamilton's equations of motion, which can be written in terms of the Poisson Brackets (PB), $\D f/ \D t = -\{f,\hat{H}\} +\partial f/\partial t$, 
where $f$ and $\hat{H}$ are functions of the canonical microscopic coordinates, $\{{\bf q}^\alpha_i\}$, and their conjugate momenta, $\{{\bm \pi}^\alpha_i\}$.
Here $\alpha$ refers to different particles and $i$ to different generalized coordinates (might be, {\it e.g.}, rotational coordinates).

We are interested in the dynamics of the coarse grained total momentum and angular momentum densities.
The microscopic fields $\hat \Phi_\mu\left(\{{\bf q}^\alpha_i\},\{{\bm \pi}^\alpha_i\},t\right)$ 
are then coarse-grained to give mesoscopic fields $\Phi_\mu(\vecr,t) = [ \hat \Phi_\mu\left(\{{\bf q}^\alpha_i\},\{{\bm \pi}^\alpha_i\},t\right)]_c$, 
and the statistical mechanics of the latter is determined by the coarse-grained Hamiltonian $H[\{\Phi_\mu\}]$. 
The reactive (non-dissipative) part of the dynamics of the coarse-grained fields is found by essentially applying the PB to these fields~\cite{Lubensky2003},
\begin{eqnarray}
\label{p1}
\frac{\partial \Phi_\mu(\vecr,t)}{\partial t} \Big|_{\rm reactive} = - \int \D\vecr' \{\Phi_\mu(\vecr),\Phi_\nu(\vecr')\}\frac{\delta H}{\delta \Phi_\nu(\vecr')} \, ,
\end{eqnarray}
where $\{\Phi_\mu(\vecr),\Phi_\nu(\vecr')\} = [ \{\hat\Phi_\mu(\vecr),\hat\Phi_\nu(\vecr')\}]_c$ and 
\begin{eqnarray}
\label{p1a}
\{\hat\Phi_\mu(\vecr),\hat\Phi_\nu(\vecr')\} = \sum_{\alpha i} \left[ \frac{\partial \hat\Phi_\mu(\vecr)}{\partial \pi_i^\alpha} \frac{\partial \hat\Phi_\nu(\vecr')}{\partial q_i^\alpha} 
- \frac{\partial \hat\Phi_\mu(\vecr)}{\partial q_i^\alpha} \frac{\partial \hat\Phi_\nu(\vecr')}{\partial \pi_i^\alpha} \right] \, .
\end{eqnarray}
Note that these equations are nothing more than a generalization of the microscopic PB when applying the chain rule and considering the fact that $\Phi_\mu(\vecr)$ are fields.
Furthermore, the PB only couples the fields $\Phi_\mu$ and $\Phi_\nu$ if they have different sign under time reversal.

However, unlike the microscopic dynamics, the PB does not give the complete mesoscopic dynamics.
The coarse-graining procedure neglects many degrees of freedom, which appears as an additional dissipative term in the coarse-grained dynamics
\begin{eqnarray}
\label{p2}
\frac{\partial \Phi_\mu(\vecr,t)}{\partial t} \Big|_{\rm dissipative} = -\int \D\vecr' \, \Gamma_{\mu\nu}(\vecr,\vecr') \frac{\delta H}{\delta \Phi_\nu(\vecr')} \, .
\end{eqnarray}
The dissipative tensor ${\bm \Gamma}$ is symmetric and positive semi definite.
It is generally a function of all fields $\{\Phi_\mu\}$, and following Curie's symmetry principle it must obey the system symmetries.
Moreover, oppositely to the reactive terms (Eq.~(\ref{p1})), 
here $\Phi_\mu$ is dissipatively coupled  to $\Phi_\nu$ only if they have the same sign under time reversal (this is the signature of dissipation).

\subsection{Poisson Brackets in our model}

The first step in calculating the PB is to determine the coarse-grained Hamiltonian. In the context of PB, which is purely mechanistic, coarse-graining refers to spatial and temporal coarse graining (average over particles is some mesoscopic volume and over microscopic timescales) rather than ensemble average, which requires discussing temperature. Here we use the simplest Hamiltonian (Eq.~(2) of the main text) that can be obtained by direct coarse-graining~\cite{tobepub}:
\begin{eqnarray}
\label{p3}
H = \int \D \vecr \, \frac{ {\bf g}^2 }{2\rho} =  \int \D \vecr\left[  \frac{ ({\bf g}^c)^2 }{2\rho}  + \frac{1}{8\rho} (\nabla\times\vecl)^2 + \vecl\cdot{\bm \omega}^c\right] 
\simeq  \int \D \vecr\left[  \frac{ ({\bf g}^c)^2 }{2\rho} + \vecl\cdot{\bm \omega}^c\right] \, ,
\end{eqnarray}
with ${\bm \omega}^c = \half \nabla \times \vecv^c$ is half the fluid vorticity.
Note that in the last equality we have omitted the term $\sim(\nabla \times \vecl)^2$ 
as it is higher order in gradients and therefore vanishes in the hydrodynamic limit (long wavelength).
We would like to find the equations of motion for both the momentum and the angular momentum densities.
Following Eq.~(\ref{p1}), the reactive parts of these equations are: 
\begin{eqnarray}
\label{p4a}
& &\dot{g}_i(\vecr) = - \int \D\vecr' \left[ \{g_i(\vecr),g_j(\vecr')\} \frac{\delta H}{\delta g_j(\vecr')} + \{g_i(\vecr),\rho(\vecr')\} \frac{\delta H}{\delta \rho(\vecr')} \right] \\
\label{p4b}
& &\dot{\ell}_i(\vecr) = - \int \D\vecr' \left[ \{\ell_i(\vecr),g_j(\vecr')\} \frac{\delta H}{\delta g_j(\vecr')} + \{\ell_i(\vecr),\rho(\vecr')\} \frac{\delta H}{\delta \rho(\vecr')} \right] \, . 
\end{eqnarray}
The fundamental PB needed to calculate these equations are related to the fields $\vecl(\vecr),\vecg^c(\vecr),\rho(\vecr)$:
\begin{eqnarray}
\label{p5}
& &\{\ell_i(\vecr),\ell_j(\vecr')\} = - \varepsilon_{ijk} \ell_k(\vecr) \delta(\vecr-\vecr') \, ,\\
\label{p5aaa}
& &\{g_i^c(\vecr),\ell_j(\vecr')\} = - \nabla'_i \left[ \ell_j(\vecr') \delta(\vecr-\vecr')\right] \, ,\\
\label{p5aa}
& &\{\ell_i(\vecr),g_j^c(\vecr')\} = \ell_i(\vecr') \nabla_j \delta(\vecr-\vecr') \, ,\\
\label{p5a}
& &\{g^c_i(\vecr),\rho(\vecr')\} = \rho(\vecr)  \nabla_i \delta(\vecr'-\vecr) \, .
\end{eqnarray}
Other PB between these fields vanish.
Note that we have assumed the system has no structural order, {\it i.e.} ${\bf K} = 0$, otherwise $\rho$ should be decomposed to $\rho^c$ and ${\bf K}$.

The equations of motion for the momentum and angular momentum are supplemented by the dynamics for the mass density
\begin{eqnarray}
\label{p4c}
& &\dot{\rho}(\vecr) = - \int \D\vecr' \left[ \{\rho(\vecr),g_j(\vecr')\} \frac{\delta H}{\delta g_j(\vecr')} + \{\rho(\vecr),\ell_j(\vecr')\} \frac{\delta H}{\delta \ell_j(\vecr')} \right]  \, . 
\end{eqnarray}
Using Eq.~(\ref{p5a}) and the fact that $\{\rho(\vecr),\ell_j(\vecr')\} = 0$ we find the usual continuity equation $\dot{\rho} + \nabla \cdot \vecg = 0$ where dissipative terms are not allowed because the mass density is conserved.

\subsubsection{Dynamics of the total momentum density}

The reactive part of the dynamics of the total momentum is then
\begin{eqnarray}
\label{p4}
\nonumber\dot{g}_i(\vecr) &=& - \int \D\vecr' \Bigg[ \Big( \{g_i^c(\vecr),g_j^c(\vecr')\}  +\{g_i^c(\vecr),g_j^s(\vecr')\}  +  \{g_i^s(\vecr),g_j^c(\vecr')\}  + \{g_i^s(\vecr),g_j^s(\vecr')\}    \Big) v_j(\vecr') \\
&-&  \{g_i(\vecr),\rho(\vecr')\} \, \frac{v^2(\vecr')} {2} \Bigg] \, .
\end{eqnarray}
The required PB for Eq.~(\ref{p4}) are
\begin{eqnarray}
\label{p6}
& &\{g_i^c(\vecr),g_j^s(\vecr')\} = -\half \varepsilon_{jkl}  \nabla'_k \nabla'_i \left[  \ell_l(\vecr') \delta(\vecr-\vecr')\right] \, ,\\
\label{p7}
& &\{g_i^s(\vecr),g_j^c(\vecr')\} = \half \varepsilon_{ikl}   \ell_l(\vecr') \nabla_k\nabla_j \delta(\vecr-\vecr') \, , \\
\label{p8}
& &\{g_i^s(\vecr),g_j^s(\vecr')\} = -\frac{1}{4} \Big( \varepsilon_{ijk}   \nabla_k\nabla'_s  \left[ \ell_s(\vecr) \delta(\vecr-\vecr') \right] 
+ \varepsilon_{ikn}\nabla_k\nabla'_n\left[\ell_j(\vecr)\delta(\vecr-\vecr')\right] \Big) \, ,\\
\label{p9}
& &\{g_i^c(\vecr),g_j^c(\vecr')\} = g_i^c(\vecr') \nabla_j \delta(\vecr-\vecr') - \nabla'_i \left[ g_j^c(\vecr') \delta(\vecr-\vecr')  \right] \, .
\end{eqnarray}
Substituting Eqs.~(\ref{p5a})-(\ref{p9}) into Eq.~(\ref{p4}) and using integration by parts (while throwing boundary terms) yields
\begin{eqnarray}
\label{p11}
\dot{g}_i = \half\rho\nabla_i\left(v^2\right) - \nabla_j \left(v_j g_i^c\right) - g_j^c\nabla_i v_j  + \half \varepsilon_{jkl} \ell_l \nabla_i\nabla_k v_j - \half \varepsilon_{ikl} \nabla_k\nabla_j\left(v_j\ell_l\right) - \frac{1}{4} \varepsilon_{ijk} \nabla_k \Big[ \ell_m \left(  \nabla_m v_j - \nabla_j v_m\right) \Big] \, .
\end{eqnarray}
The form of Eq.~(\ref{p11}) is not satisfactory.
Since the momentum is conserved, one expects to be able to write $\dot{g}_i +  \nabla_j \left(v_j g_i\right) = \nabla_j \sigma_{ij}$.
The resolution lies in expressing $\vecg^c$ in terms of $\vecg$ and $\vecl$ using Eq.~(\ref{m7a}).
Rewriting the first three terms of the right-hand-side of Eq.~(\ref{p11}), we have 
\begin{eqnarray}
\label{p12}
\half\rho\nabla_i\left(v^2\right) - \nabla_j \left(v_j g_i^c\right) - g_j^c\nabla_i v_j  = - \nabla_j \left(v_j g_i\right) + \half\varepsilon_{imn} \nabla_j \left(v_j\nabla_m\ell_n\right) 
+\half \varepsilon_{jmn} (\nabla_m\ell_n)(\nabla_i v_j) \, ,
\end{eqnarray}
and 
\begin{eqnarray}
\label{p13}
& &\dot{g}_i +  \nabla_j \left(v_j g_i\right) = \nabla_j \sigma_{ij} \, ,\\
\label{p13a}
& & \sigma_{ij} =  -\half \left(  \varepsilon_{jkn}\delta_{il}\ell_n +  \varepsilon_{iln}\delta_{jk}\ell_n + \half \left[\varepsilon_{ijl}\ell_k  
- \varepsilon_{ijk}\ell_l \right] \right) \nabla_l v_k  \, .
\end{eqnarray}
Although it is not clear from the structure above, it can be verified that the stress tensor is symmetric.
The antisymmetric part of the stress tensor is $\sigma^a_{ij} = \half( \sigma_{ij} - \sigma_{ji})$, and using Eq.~(\ref{p13a}) we have
\begin{eqnarray}
\label{p13b}
\nonumber\sigma_{ij}^a &=&   -\frac{1}{4} \Big[ \ell_n \left( \varepsilon_{jkn}\delta_{il}   -  \varepsilon_{ikn}\delta_{jl}  
+  \varepsilon_{iln}\delta_{jk} - \varepsilon_{jln}\delta_{ik} \right)  + \varepsilon_{ijl}\ell_k  - \varepsilon_{ijk}\ell_l  \Big] \nabla_l v_k  \\
\nonumber&=&  -\frac{1}{4}  \ell_n  \Big[  \varepsilon_{jln}  \left( \nabla_i v_l - \nabla_l v_i\right)   -  \varepsilon_{iln} \left( \nabla_j v_l - \nabla_l v_j \right)  
+ \varepsilon_{ijl} \left( \nabla_l v_n - \nabla_n v_l \right)    \Big] \\
\nonumber&=&  -\frac{1}{4}  \ell_n  \varepsilon_{mst}  \Big[  \varepsilon_{jln} \varepsilon_{ilm}   -  \varepsilon_{iln} \varepsilon_{jlm}
+ \varepsilon_{ijk} \varepsilon_{knm}    \Big] \nabla_s v_t \\
&=& -\frac{1}{4}  \ell_n  \Big[ \delta_{ij}\delta_{nm} - \delta_{jm}\delta_{in} - \delta_{ij}\delta_{nm} + \delta_{im}\delta_{nj} + \delta_{in}\delta_{jm} - \delta_{im}\delta_{jn}  \Big] \nabla_s v_t = 0 \, , 
\end{eqnarray}

The stress tensor can then be written in a more familiar form:
%
\begin{eqnarray}
\label{p14}
\sigma_{ij} = \left[  -\frac{1}{4} \ell_n \left(  \varepsilon_{jln}\delta_{ik} + \varepsilon_{iln}\delta_{jk} 
+  \varepsilon_{ikn}\delta_{jl} +  \varepsilon_{jkn}\delta_{il} \right) \right]  \nabla_l v_k  \, \equiv \eta^o_{ijkl} \nabla_l v_k \, .
\end{eqnarray}
%
This form is the same as was phenomenologically written for plasma with magnetic fields~\cite{LLfluids}, where $\vecl$ plays the role of a magnetic field.
Since in deriving Eq.~(\ref{p14}) we did not assume anything about $\vecl$ it can be any function of space ans time.
To compare with recent works of active spinning particles we also write the stress tensor in 2D:
\begin{eqnarray}
\label{p15}
\sigma_{ij} =   -\frac{\ell}{2} \Big(  \nabla_i \varepsilon_{jm} v_m +  \varepsilon_{ik} \nabla_k v_j  \Big)  \, .
\end{eqnarray}
This is very similar to the expression for the `odd' part of the stress tensor in Ref.~\cite{Banerjee2017}.
There is, however, a crucial difference between the two formulations. Here the stress tensor is that for the dynamics of the total momentum density while that of Ref.~\cite{Banerjee2017} is that of the CM momentum density.
This is crucial (as explained in the main text) as in our formalism the stress tensor must be symmetric due to conservation of the total momentum~\footnote{In the presence of external or active torques there is an antisymmetric stress: $\half\varepsilon_{ijk}\tau_k$. There are no other antisymmetric stresses.},
which also constraints the form of the dissipation (see again discussion in the main text). 
Our results, Eqs.~(\ref{p14})-(\ref{p15}) are valid for any $\ell$, constant or not.

The dissipative terms can be obtained using Eq.~(\ref{p2}), and are written below in Eq.~(\ref{g2}).

\subsubsection{Dynamics of angular momentum density}

It is straightforward to compute the PB of Eq.~(\ref{p4b}) using Eqs.~(\ref{p5})-(\ref{p5aa}) and $\{\ell_i(\vecr),\rho(\vecr')\} = 0$ (because we assume ${\bf K} = 0$).
The reactive part of the $\vecl$ dynamics is then
\begin{eqnarray}
\label{l1}
\dot{\ell}_i(\vecr) + \nabla_j\left(\ell_i v_j\right) = \varepsilon_{ijk}\ell_k\omega_j \, .
\end{eqnarray}
In 2D, the right-hand-side vanishes. 
Dissipation in $\vecl$ is constrained by the requirement that for uniform rotations of the system, in which ${\bm \Omega}={\bm \omega}^c$,
there is no dissipation.
Thus, the dissipative term must has the form $\Gamma_{ij} (\Omega_j - \omega^c_j)$~\cite{Lubensky2005,NJP}.
The form of ${\bm \Gamma}$ is constraint by symmetry (see Section~\ref{symmetry}) and the general requirements of the dissipative tensor (see paragraph below Eq.~(\ref{p2})).
It is thus written as
\begin{eqnarray}
\label{l2}
\Gamma_{ij} = g_1 \delta_{ij} + g_2 \ell_i \ell_j \, .
\end{eqnarray}
In the main text we assume the dissipation to be isotropic by choosing $g_1 = \Gamma$ and $g_2=0$ such that $\Gamma_{ij} = \Gamma \delta_{ij}$. The next step is to express ${\bm \omega}^c$ in terms of ${\bm \omega}$ and $\vecl$:
\begin{eqnarray}
\label{l3}
\nonumber\dot{\ell}_i(\vecr) + \nabla_j\left(\ell_i v_j\right) &=&  \varepsilon_{ijk}\ell_k\omega_j - \Gamma \left(\Omega_i - \omega_i\right) 
-\frac{\Gamma}{4}\nabla_j\left[ \frac{1}{\rho}(\nabla_i\ell_j - \nabla_j\ell_i)  \right] +\tau_i \\
&\simeq& \varepsilon_{ijk}\ell_k\omega_j - \Gamma \left(\Omega_i - \omega_i\right) +\tau_i \, ,
\end{eqnarray}
where we have added an external (or active) torque ${\bm \tau}$ and in the second equality we have neglected terms $\sim\nabla^2\vecl$.
This is Eq.~(7) of the main text.

The steady state of the bulk fluid for ${\bm \tau} = 0$ is ${\bf v} = 0$ (assuming no other external forcing). For non-zero  ${\bm \tau}$ we write ${\bm \Omega} = {\bm \Omega}^0 + \delta{\bm \Omega}$ and assume ${\bm \nabla}{\bf v}$ (and thus also ${\bm \omega}$) scales as $ \delta{\bm \Omega}$. The long-wavelength low-frequency solution is ${\bm \Omega}_0 \simeq {\bm \tau}/\Gamma$ and $\vecl_0 = {\bf I} \cdot {\bm \tau}/\Gamma$.
This effectively creates a constant (in time) angular momentum density, which is related to the odd viscosity coefficient (Eq.~(5) in the main text).

Note that in order to obtain the dissipative term in the dynamics of $\vecl$ from Eq.~(\ref{p2}), 
the Hamiltonian should have included also a term $\half \vecl \cdot {\bf I}^{-1} \cdot \vecl$ as in Refs.~\cite{Lubensky2005,NJP}.
We show elsewhere~\cite{tobepub} that such a term indeed exists in the Hamiltonian alongside the standard $\vecg^2/(2\rho)$ term we consider in this work.
However, the addition of such a term in the Hamiltonian has no effect on the reactive part of the dynamics for structurally isotropic materials ($I_{ij}\sim \delta_{ij}$)~\cite{tobepub}.

\subsubsection{Poisson-Brackets for the pressure}
\label{sec:PB_pressure}
In order to obtain the pressure that is added in Eq.~(6) of the main text the Hamiltonian of Eq.~(\ref{p3}) must be supplemented by a free-energy $F[\rho] = \int\D\vecr f\left(\rho,\nabla\rho\right)$:
\begin{eqnarray}
\label{pr1}
H = \int \D \vecr \, \frac{ {\bf g}^2 }{2\rho} + F[\rho(\vecr)]  \, .
\end{eqnarray}
Since  $\{\ell_i(\vecr),\rho(\vecr')\} = 0$ this addition does not affect the angular momentum dynamics, but the dynamics for the total momentum is modified
\begin{eqnarray}
\label{pr2}
\nonumber\dot{g}_i(\vecr) &=& - \int \D\vecr' \Bigg[ \Big( \{g_i^c(\vecr),g_j^c(\vecr')\}  +\{g_i^c(\vecr),g_j^s(\vecr')\}  +  \{g_i^s(\vecr),g_j^c(\vecr')\}  + \{g_i^s(\vecr),g_j^s(\vecr')\}    \Big) v_j(\vecr') \\
&+&  \{g_i(\vecr),\rho(\vecr')\} \left( \frac{\delta F}{\delta\rho(\vecr')} - \frac{v^2(\vecr')} {2} \right)\Bigg] \, ,
\end{eqnarray}
such that with the use of Eq.~(\ref{p5a}) the resulting stress tensor of Eq.~(\ref{p14}) becomes $\sigma_{ij} = \eta^o_{ijkl} \nabla_l v_k - P\delta_{ij}$ where $P = \rho\frac{\delta F}{\delta\rho} - f$. To obtain this we have also used the identity $\rho \nabla_i \frac{\delta F}{\delta\rho} = \nabla_i P$.

\section{Phenomenological derivation of the stress tensor -  general symmetry considerations} \label{symmetry}

Using the Hamiltonian of Eq.~(\ref{p3}) as the free-energy, $F$ (assuming local equilibrium), 
and writing a general conservation equation for the total momentum,  $\dot{g}_i + \nabla_j(v_jg_i)= \nabla_j \sigma_{ij}$,
one finds the thermodynamic entropy production rate to be (see {\it e.g.} Ref.~\cite{NJP} Appendix~B or~\cite{Mazur}),
\begin{eqnarray}
\label{g1}
T\frac{\D S}{\D t} =  \int\D\vecr\, u_{ij} \sigma^k_{ij} \, ,
\end{eqnarray}
where $u_{ij} = \half(\nabla_i v_j + \nabla_j v_i)$ and $\sigma_{ij}^k$ is the kinetic part of the stress tensor ({\it i.e.}, does not include elastic contributions).
Within the framework of linear irreversible thermodynamics the thermodynamic fluxes ($\sigma_{ij}^k$)  are expanded to linear order in the thermodynamic forces ($u_{ij}$).
The coefficients in this expansion are called the kinetic coefficients, and they must obey (i) the system symmetries and (ii) Onsager relations.
The first condition implies that the kinetic coefficients are written in terms of the system order parameters, which is $\vecl$ in our case. Notably, this means that a kinetic coefficient that has a part with odd number of ${\bm \ell}$'s will behave differently under time reversal than other order-parameter dependent parts (similar to what happens when the order parameter is magnetization).

Specifically, introducing the viscosity tensor, $\sigma^k_{ij} = \eta_{ijkl} \nabla_l v_k$, and writing $\eta_{ijkl}$ in terms of $\delta_{ij}$, $\varepsilon_{ijk}$, and ${\bm \ell}$  
we have $\eta_{ijkl} = \eta_{ijkl}^e + \eta_{ijkl}^o$  with 
\begin{eqnarray}
\label{g2}
\nonumber \eta^e_{ijkl} &=& \zeta \delta_{\alpha\beta}\delta_{\gamma\delta} + \eta\left( \delta_{\alpha\gamma}\delta_{\beta\delta} + \delta_{\alpha\delta}\delta_{\beta\gamma} \right) + \alpha_1\left( \delta_{\alpha\beta}\ell_\gamma \ell_\delta + \delta_{\gamma\delta}\ell_\alpha \ell_\beta \right) 
+ \alpha_2\Big( \ell_\alpha \ell_\gamma \delta_{\beta\delta} + \ell_\beta \ell_\gamma \delta_{\alpha\delta} + \ell_\alpha \ell_\delta \delta_{\beta\gamma} + \ell_\beta \ell_\delta \delta_{\alpha\gamma} \Big) \\
&+& \alpha_3 \ell_\alpha \ell_\beta \ell_\gamma \ell_\delta \, ,\\
\label{g3}
\eta^o_{ijkl} &=& \beta_1  \left(  \varepsilon_{jln}\delta_{ik} + \varepsilon_{iln}\delta_{jk} 
+  \varepsilon_{ikn}\delta_{jl} +  \varepsilon_{jkn}\delta_{il} \right) \ell_n + \beta_2  \left(  \varepsilon_{jln} \ell_i \ell_k  + \varepsilon_{iln}\ell_j \ell_k 
+  \varepsilon_{ikn}\ell_j \ell_l  +  \varepsilon_{jkn}\ell_i \ell_l  \right) \ell_n \, ,
\end{eqnarray}
such that $\eta^e_{ijkl} = \eta^e_{klij}$ and $\eta^o_{ijkl} = -\eta^o_{klij}$.
All of the coefficients above are generally a function of $\vecl^2$.
Indeed, the time reversal signatures of ${\bm \eta}^o$ and ${\bm \eta}^e$ are different.
The former does not contribute to the entropy production rate, hence, it is non-dissipative or reactive, and only ${\bm \eta}^e$ is dissipative.
These general symmetry arguments allow for terms in ${\bm \eta}^o$ that are not observed using our microscopic model.
We expect such terms to appear when mode-coupling terms, not considered in our  current treatment, are included.
The $\vecl$-dependent terms in ${\bm \eta}^e$ are neglected in the main text.

\section{Odd viscosity from the center-of-mass momentum} \label{CM}

In order to find the reactive part of the CM momentum we calculate
\begin{eqnarray}
\label{cm1}
\dot{g}^c_i(\vecr) &=& - \int \D\vecr' \left[ \{g_i^c(\vecr),g^c_j(\vecr')\} \frac{\delta H}{\delta g^c_j(\vecr')} + \{g_i^c(\vecr),\ell_j(\vecr')\} \frac{\delta H}{\delta \ell_j(\vecr')} + \{g_i^c(\vecr),\rho(\vecr')\} \frac{\delta H}{\delta \rho(\vecr')} \right] \, .
\end{eqnarray}
The required PB are given in Eqs.~(\ref{p5aaa}), (\ref{p5a}), and (\ref{p9}).
Using these PB Eq.~(\ref{cm1}) reads,
\begin{eqnarray}
\label{cm3}
\dot{g}^c_i +  \nabla_j \left(v^c_j g_i^c\right)  = 
- \half \nabla_j \left[ v_i^c (\nabla \times \vecl)_j + \varepsilon_{jmn} \ell_n\nabla_i v_m^c   \right] \, .
\end{eqnarray}
If $\vecl$ is constant in space this expression simplifies greatly,
\begin{eqnarray}
\label{cm4}
\dot{g}^c_i +  \nabla_j \left(v^c_j g_i^c\right)  = 
- \half \nabla_j \left[  \varepsilon_{jmn} \ell_n\nabla_i v_m^c   \right] \, ,
\end{eqnarray}
which becomes in 2D 
\begin{eqnarray}
\label{cm5}
\dot{g}^c_i +  \nabla_j \left(v^c_j g_i^c\right)  = 
- \frac{\ell}{2} \nabla_j \left[  \varepsilon_{jm}  \nabla_i v_m^c  \right] \, .
\end{eqnarray}
The term in the right-hand-side is the same (except that here $\vecv \to \vecv^c$) as the first `odd' term in the dynamics of the total momentum, Eq.~(\ref{p15}).

The symmetric dissipative term is the same as for the total velocity (with ${\bf v} \to {\bf v}^c$ of course), see Eq.~(\ref{g2}).
However, for the CM momentum the stress tensor is not necessarily symmetric.
In fact, conservation of total angular momentum dictates that the (internal) body term in the dynamics of $\vecl$ must be equal to $-\varepsilon_{ijk}\sigma_{jk}$~\cite{NJP}
\footnote{Strictly speaking, it is equal to the part of $-\varepsilon_{ijk}\sigma_{jk}$ that is not a divergence of any tensor.}.
Therefore, the dissipative antisymmetric term is $\frac{\Gamma}{2}  \varepsilon_{ijk}  \nabla_j \left(  \Omega_k - \omega^c_k  \right)$.
One can then write the full equation of motion for ${\bf g}^c$ as $\dot{g}^c_i +  \nabla_j \left(v^c_j g_i^c\right)  = \nabla_j \sigma_{ij}^c$ with
\begin{eqnarray}
\label{cm6}
\sigma_{ij}^c =   - P\delta_{ij} + \eta_{ijkl}(\nabla_l v^c_{k}) - \half v_i^c (\nabla \times \vecl)_j 
+\frac{\Gamma}{2}  \varepsilon_{ijk}  \left(  \Omega_k - \omega^c_k  \right)\, .
\end{eqnarray}
and where 
\begin{eqnarray}
\label{cm7}
\eta_{ijkl} = - \half  \varepsilon_{jkn}\delta_{il}  \ell_n + \eta^e_{ijkl} \, .
\end{eqnarray}
Note that we have also added the pressure term that can be derived from PB if one includes a density dependent free-energy (see Sec.~\ref{sec:PB_pressure} and \cite{Lubensky2003}).
The dynamics of $\vecl$ in term of $\vecv^c$ is found by substituting Eq.~(\ref{m7a}) in Eq.~(\ref{l3}):
\begin{eqnarray}
\label{cm9}
\dot{\ell}_i(\vecr) + \nabla_j\left(\ell_i v^c_j\right) \simeq  \varepsilon_{ijk}\ell_k\omega^c_j - \Gamma \left(\Omega_i - \omega^c_i\right)  + \tau_i \, ,
\end{eqnarray}
where, as before, we  neglect terms $\sim\nabla^2\vecl$.
We now have all the dynamics in terms of the CM momentum.
In this representation not all `odd'  terms are present, but instead there is an antisymmetric dissipative term connecting the CM momentum and the angular momentum densities.

Lets us show that the dynamics of the total momentum  (Eq.~(4) of the main text) can be recovered from the CM dynamics. To do so we start by substituting Eq.~(\ref{cm9}) into Eq.~(\ref{cm6}) yielding,
\begin{eqnarray}
\label{cm9a}
\sigma_{ij}^c  =  -P\delta_{ij} + \left( \eta^e_{ijkl} + \eta^o_{ijkl}\right)  \nabla_l v^c_k - \half \left[ v_i^c (\nabla \times \vecl)_j + v_j^c (\nabla \times \vecl)_i  \right]
+  \half \varepsilon_{ijk} \left(\tau_k - \dot{\ell}_k\right)   \, ,
\end{eqnarray}
with $\eta^o_{ijkl}$ as defined in Eq.~(5) of the main text. This stress has the complete odd viscosity but also two terms that do not appear in the total momentum formulation (see Eq.~(4) of the main text and below): a term that depends on $\nabla\times\vecl$ and an antisymmetric term that depends on the rate of change of the internal angular momentum. The former seems to break  Galilean invariance, however, when the Galilean transformation is applied to all fields including the angular momentum one can verify that this term is Galilean invariant as required. The dynamics of the total angular momentum can be readily recovered from Eq.~\eqref{cm9a} by using the definition of the total momentum~(\ref{m7a}), which implies that the total momentum obeys $\dot{g}_i + \nabla_j(v_j g_i) = \nabla_j \sigma_{ij}$ with $\sigma_{ij} = \sigma^c_{ij} + \sigma^\ell_{ij}$ and
\begin{eqnarray}
\label{cm9aa}
\nabla_j \sigma_{ij}^\ell = \half \varepsilon_{ijk} \nabla_j \dot{\ell}_k + \nabla_j\left( v_j g_i - v_j^c g_i^c  \right)  \, .
\end{eqnarray}
Adding Eqs.~\eqref{cm9aa} and Eq.~\eqref{cm9a} gives the desired total momentum stress tensor.

It is tempting to eliminate ${\bm \Omega}$ from the CM dynamics, Eq.~\eqref{cm6}, with the hope that the antisymmetric dissipative stress will be replaced by the missing odd terms. Below we show that this is not the case and that the proper odd viscosity is only obtained for the total momentum. We eliminate ${\bm \Omega}$ by solving Eq.~(\ref{cm9}) as in the solution in the total momentum formulation in which we take the long-wavelength low-frequency limit and write ${\bm \Omega} = {\bm \Omega}^0 + \delta{\bm \Omega}$, $\rho = \rho_0 + \delta\rho$ while assuming $\delta{\bm \Omega}$ and $\delta\rho$ are of the order of ${\bm\nabla}{\bf v}^c$. The result is
\begin{eqnarray}
\label{cm10}
\nonumber& &\Omega_i^0 =  \tau_i/\Gamma \, , \\
& & \delta\Omega_i = \omega_i^c + \frac{1}{\Gamma}\left[  \half\ell_k^0 \left(   \nabla_k v^c_i - \nabla_i v^c_k  \right) - \rho_0  {\bf v}^c\cdot\nabla \left(\tilde{I}\Omega_i^0\right)  \right] \, ,
\end{eqnarray}
which with the aid of the continuity equation yields
\begin{eqnarray}
\label{cm10a}
\dot{\ell}_i \simeq -\ell^0_i \nabla\cdot\vecv - \tilde{I}\Omega^0_i \vecv\cdot\nabla\rho_0  \, ,
\end{eqnarray}
where $\vecl^0 = {\bf I} \cdot {\bm \Omega}^{0} = \rho_0 \tilde{I} {\bm \Omega}^{0}$. This solution is valid for any ${\bm \tau}$, even if it is inhomogeneous and not constant in time. Substituting Eq.~(\ref{cm10a}) into Eq.~(\ref{cm9a})  gives the low-frequency long-wavelength dynamics of the CM momentum after relaxation of ${\bm \Omega}$. The CM stress tensor can then be written as:
\begin{eqnarray}
\label{e27a}
\nonumber \sigma_{ij}^c &=&  \left(\eta^e_{ijkl}+ \eta^o_{ijkl} \right) \nabla_l v^c_k
- P\delta_{ij} -   \half  \left( \varepsilon_{imn} v^c_j   + \varepsilon_{jmn} v^c_i  \right) \left(\nabla_m\ell^0_n\right) \\
\nonumber&+& \half\varepsilon_{ijn} \left( \ell^0_n \nabla\cdot\vecv^c + \tilde{I}\Omega_n^0 \vecv^c\cdot\nabla\rho_0 \right) + \frac{1}{2} \varepsilon_{ijn} \tau_n  \\
&=&  \left(\eta^e_{ijkl}+ \tilde\eta^o_{ijkl} \right) \nabla_l v^c_k
- P\delta_{ij} -   \half  \left( \varepsilon_{imn} v^c_j   + \varepsilon_{jmn} v^c_i  \right) \left(\nabla_m\ell^0_n\right) 
+ \half\varepsilon_{ijn} \left( \tilde{I}\Omega_n^0 \vecv^c\cdot\nabla\rho_0 + \tau_n \right) \, ,
\end{eqnarray}
where
\begin{eqnarray}
\label{e27ca}
\tilde\eta^o_{ijkl} = -\frac{1}{4} \ell^0_n \left(  \varepsilon_{jln}\delta_{ik} + \varepsilon_{iln}\delta_{jk} +  \varepsilon_{ikn}\delta_{jl} +  \varepsilon_{jkn}\delta_{il} - 2\varepsilon_{ijn}\delta_{kl} \right) \, ,
\end{eqnarray}
is the CM odd viscosity, which clearly does not obey Onsager's reciprocity relations. This is most easily seen when considering both $\rho_0$ and $\Omega^0$ to be constant such that 
\begin{eqnarray}
\label{e27c}
\sigma_{ij}^c =  \left(\eta^e_{ijkl}+ \tilde\eta^o_{ijkl} \right) \nabla_l v^c_k - P\delta_{ij} \, .
\end{eqnarray}

Moreover, even after relaxation of ${\bm \Omega}$, the CM stress tensor is not symmetric and its antisymmetric contribution is not simply $\half\varepsilon_{ijk} \tau_k$, but rather contains other contributions that originate in $\dot{\vecl}$. Then, if one disregards $\dot{\vecl}$ the balance of angular momentum is not obeyed. This leads to the remarkable conclusion that the CM formulation is generally insufficient to describe the system dynamics when $\vecl$ is non-vanishing. A specific situation in which the CM formulation does provide a proper description of the system dynamics is when both ${\bm \Omega}^0$ and $\rho_0$ are constants and the fluid is incompressible (in this case $\dot{\vecl}=0$) such that  $\sigma_{ij}^c =  \left(\eta^e_{ijkl}+ \eta^o_{ijkl} \right) \nabla_l v^c_k - P\delta_{ij}$. Note that even in this very specific case, in which the complete odd viscosity appears in the CM stress tensor, $\sigma_{ij}^c$ is symmetric and does not contain the antisymmetric dissipative term.

\subsection{External torque includes bulk friction} \label{seq:bulk_friction}

In many cases, the torque in Eq.~\eqref{cm9} contains friction terms, which in their simplest form are $-\Gamma^\Omega {\bm \Omega}$. Such a friction term is more common in 2D odd fluids in which the molecules move on a surface, but it may also appear in 3D odd fluids when the solvent degrees of freedom are neglected or when molecules move within a porous medium. In this case we have ${\bm \tau} = \tilde{\bm \tau} - \Gamma^\Omega {\bm \Omega}$ where $\tilde{\bm \tau}$ is a result of, {\it e.g.}, an external field. Eliminating ${\bm \Omega}$ can still be done straightforwardly using the same method as described above where now the solution for $\bm\Omega$ in the long-wavelength long-time limit is
\begin{eqnarray}
\label{cmf1}
\nonumber& &\Omega_i^0 =  \frac{1}{\Gamma + \Gamma^\Omega}\, \tilde\tau_i\, , \\
& & \delta\Omega_i = \frac{\Gamma}{\Gamma + \Gamma^\Omega} \, \omega_i^c + \frac{1}{\Gamma+ \Gamma^\Omega}\left[  \half\ell_k^0 \left(   \nabla_k v^c_i - \nabla_i v^c_k  \right) - \rho_0  {\bf v}^c\cdot\nabla \left(\tilde{I}\Omega_i^0\right)  \right] \, ,
\end{eqnarray}
instead of Eq.~\eqref{cm10}. The CM stress tensor after elimination of ${\bm\Omega}$ has the same form as in Eq.~(\ref{e27a}) with ${\bm\tau}\to\tilde{\bm \tau}$ and $\vecl^0=\rho\tilde{I}{\bm\Omega}^0$ where ${\bm\Omega}$ is now written in Eq.~\eqref{cmf1}. The conclusions from the discussion in Sec.~\ref{CM}, where $\Gamma^\Omega=0$, thus remains valid.

\begin{figure}
	\begin{centering}
		\includegraphics[width=0.5\textwidth]{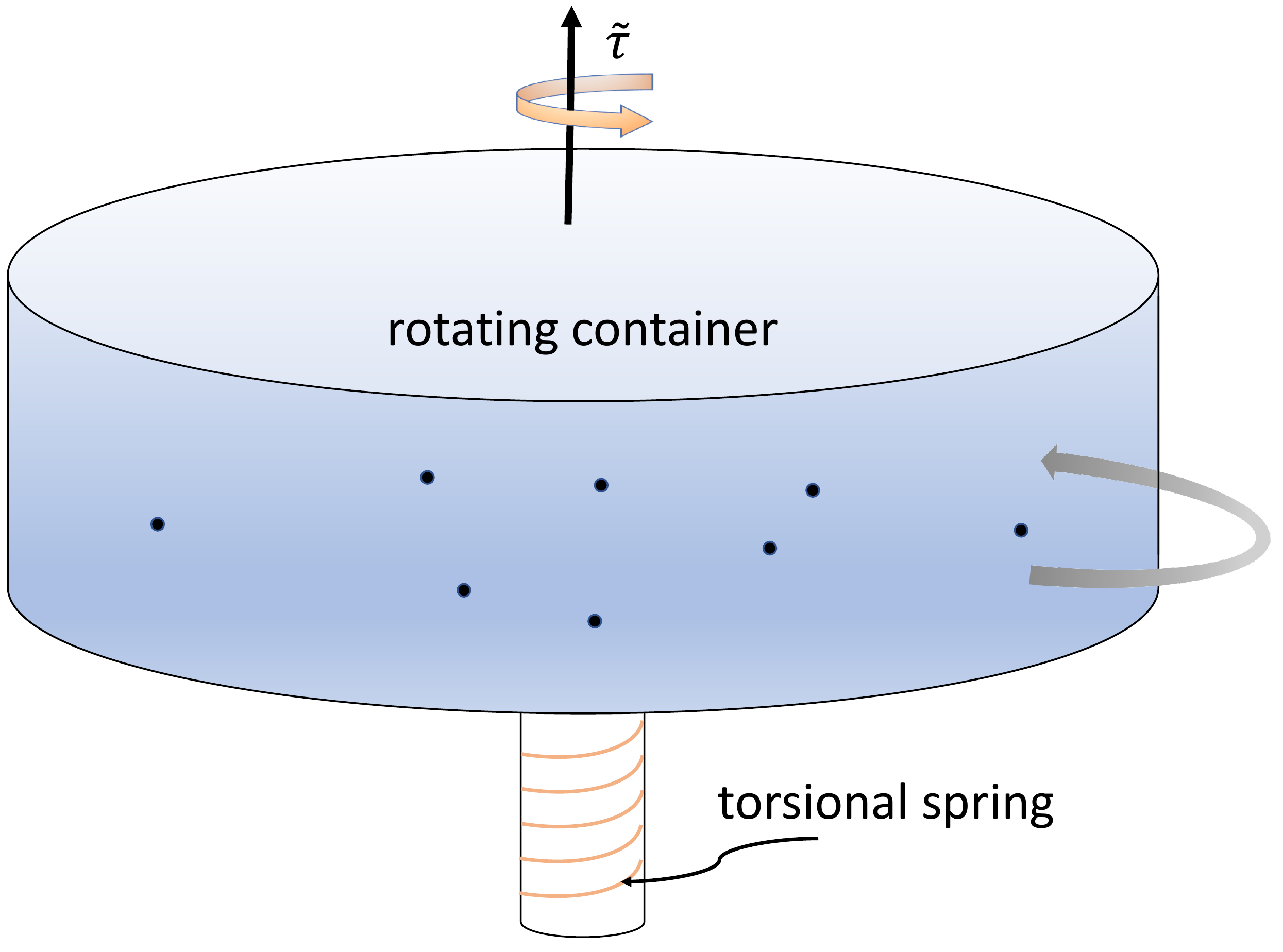}
		\par\end{centering}
	\caption{ Illustration of a rotating system that contains particles (small black circles)  on which an external body torque $\tilde{\bm \tau}$ operates. The container boundaries exerts torques on the particles $-\Gamma^s\left({\bm \Omega}-\dot{\bm \theta}\right)$ where ${\bm \theta}$ is the twist angle of the torsional spring that has a very large spring constant $\kappa$.
	}
	\label{fig:rotateing_system} 
\end{figure}

\subsection{Steady state solutions}

It is important in the presence of body torques (be external or active) to verify that the long-wavelength long-time solutions we find correspond to steady-state solutions, where the total angular momentum ${\bf L} = \int_V \left[\vecr\times\vecg^c +\vecl\right]$ does not grow in time, $\D {\bf L}/ \D t = {\bf T}^{\rm tot}= 0$, where ${\bf T}^{\rm tot}$ is the total torques on the system. An approximate way of achieving this is by considering bulk friction as in Sec.~\ref{seq:bulk_friction} such that ${\bf T}^{\rm tot} = \int_V \left(\tilde\tau - \Gamma^\Omega{\bm\Omega}\right)$ and taking the limit $\Gamma^\Omega \gg \Gamma$, which implies that ${\bm \Omega} \simeq \tilde{\bm \tau}/\Gamma^\Omega$~\cite{Banerjee2017}. Conservation of angular momentum in this case is valid (approximately) only for small enough systems because $\D {\bf L}/ \D t \sim V\Gamma/\Gamma^\Omega$. 

A more rigorous explanation lies in the system boundaries. We imagine the system being attached to an immobile surface with a torsional spring with very large spring constant $\kappa$, see Fig.~\ref{fig:rotateing_system}. The system boundaries exerts torques on the particles $-\Gamma^s\left({\bm \Omega}-\dot{\bm \theta}\right)$ where $\bm \theta$ is the twist angle of the torsional spring. By Newton's third law, the particles exerts an equal and opposite torque on the system boundaries so that the torque on the container is
\begin{eqnarray}
\label{ss1}
{\bf T}^b = -\kappa{\bm \theta} + \oint_S \Gamma^s\left({\bm \Omega}-\dot{\bm \theta}\right) \, ,
\end{eqnarray}
while the torque on the particles is
\begin{eqnarray}
\label{ss2}
{\bf T}^p = \int_V \left(\tilde{\bm \tau} - \Gamma^\Omega{\bm \Omega}\right) - \oint_S \Gamma^s\left({\bm \Omega}-\dot{\bm \theta}\right) \, ,
\end{eqnarray}
such that the total torque on the system is ${\bf T}^{\rm tot} = {\bf T}^p + {\bf T}^b = -\kappa{\bm \theta} + \int_V \left(\tilde{\bm \tau} - \Gamma^\Omega{\bm \Omega}\right)$. Conservation of angular momentum requires that ${\bf T}^{\rm tot}=0$, namely 
\begin{eqnarray}
\label{ss3}
{\bm \theta} = \frac{1}{\kappa} \int_V \left(\tilde{\bm \tau} - \Gamma^\Omega{\bm \Omega}\right) \, .
\end{eqnarray}
Because in practice $\kappa \to \infty$ the rotation angle of the container ${\bm \theta} \to 0$. Note that conservation of angular momentum does not impose any constraints on $\Gamma$, $\Gamma^\Omega$ or $\Gamma^s$.

Importantly, the system boundaries serve as an angular momentum sink (due to the large size of $\kappa$) allowing for steady-state solutions for any value of $\tilde{\bm\tau}$. Specifically, it is readily deduced from Eq.~\eqref{cmf1}  (or Eq.~\eqref{e27a}) that a constant $\tilde{\bm\tau}$ produce (approximately) a constant angular velocity of the spinning particles, that can obtain any value (it only depends on the friction coefficients and the body torque $\tilde{\bm\tau}$).

\section{Internal active driving}

\begin{figure}
	\begin{centering}
		\includegraphics[width=0.7\textwidth]{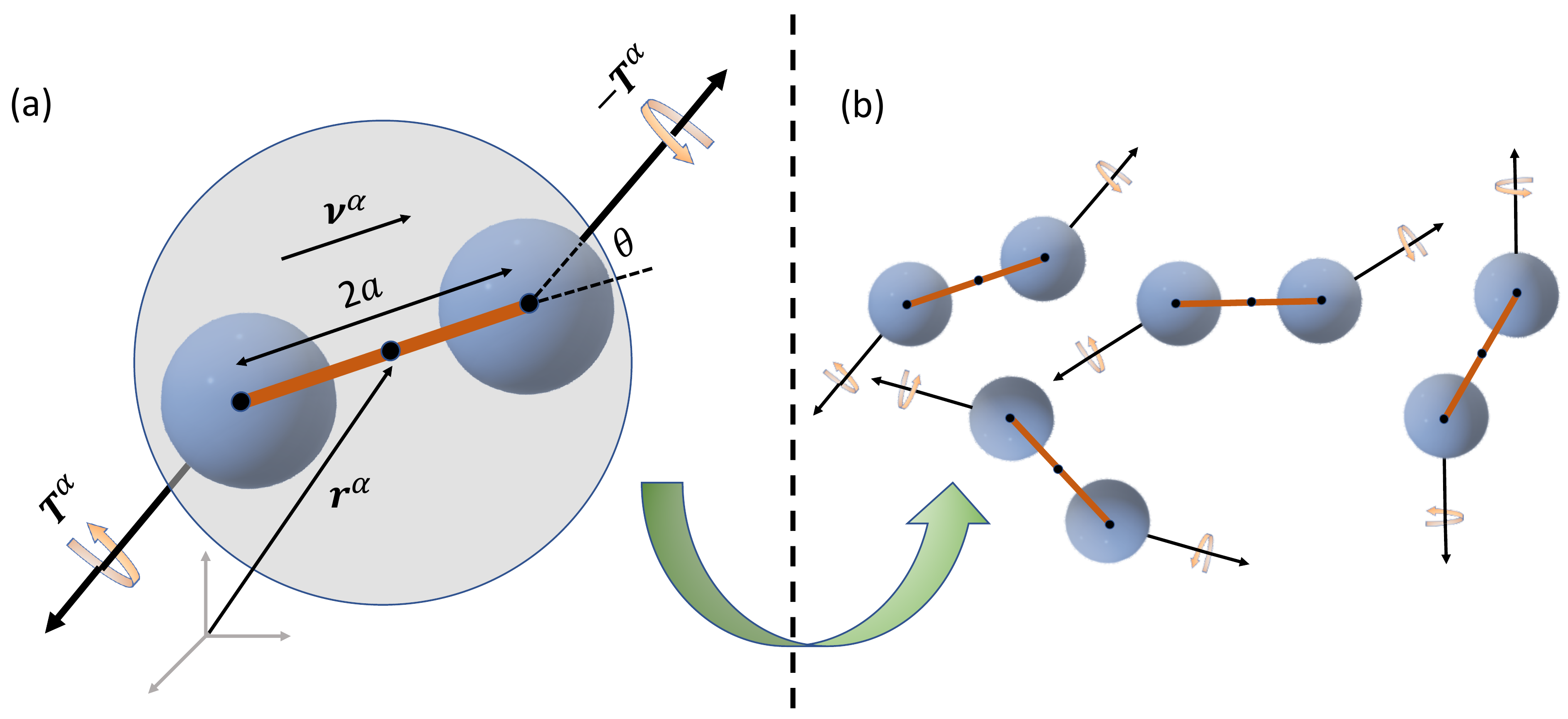}
		\par\end{centering}
	\caption{ (a) Illustration of an active torque dipole. 
		A particle whose CM located at $\vecr^\alpha$ exerts two opposite torque monopoles $\mp{\bf T}^\alpha$ at $\vecr^\alpha \pm {\bf a}^\alpha/2$.
		The two opposite torques are separated by distance $a$, hence creating a torque dipole $\sim {\bf T}^\alpha a$.		
		In (b) a random collection of such torque dipoles create a torque-dipole density $\sim\nabla n$ with $n$ being the particles' density.
	}
	\label{fig:dipole} 
\end{figure}

Internal active torques are abundant in biology and are present whenever activity is not due to an external field.
Such internal torques must be a divergence of something such that the total active torque vanishes.
For an isotropic system (say an isotropic active gel) the active torque $\sim \nabla n$, where $n$ is the density of torque dipoles. 
This corresponds to randomly oriented torque dipoles and can be seen from a simple argument.
Consider a particle with CM position $\vecr^\alpha$ that exerts two opposite torque monopoles $\mp{\bf T}^\alpha$ at $\vecr^\alpha \pm {\bf a}^\alpha/2$, 
respectively (see Fig.~\ref{fig:dipole}(a)).
Such particle creates a torque dipole density
\begin{eqnarray}
\label{t2}
{\bm \tau}^\alpha = {\bf T}^\alpha  \left[ \delta\left(\vecr - \vecr^\alpha  + {\bf a}^\alpha /2 \right) - \delta\left(\vecr - \vecr^\alpha  - {\bf a}^\alpha /2 \right) \right] 
\simeq  {\bf T}^\alpha  \left( \boldsymbol{a}^\alpha  \cdot \nabla \right) \delta\left(\vecr - \vecr^\alpha  \right) \, .
\end{eqnarray}
Averaging over ensemble of such torque dipoles (mesoscopic average) with ${\bf  a}^\alpha  = a {\bm\nu}^\alpha $ 
and assuming ${\bf T}^\alpha  = T {\bm\xi}^\alpha$ (here both ${\bm\nu}^\alpha$ and ${\bm\xi}^\alpha$ are unit vectors and the angle between them is $\theta$, which is the same for all particles) we get for isotropically distributed torque dipoles,
\begin{eqnarray}
\label{t2a}
\tau   = T a \, \nabla \cdot\left<\sum_\alpha {\bm \nu}^\alpha  {\bm \xi}^\alpha  \delta(\vecr-\vecr^\alpha )\right>_{\rm meso} 
= \frac{Ta}{d} \left(\cos\theta\right) \nabla  n(\vecr) \, ,
\end{eqnarray}
where $d$ is the dimensionality.
Here we have used the fact that for an isotropic system any second-rank tensor $\sim \delta_{ij}$.
Then by writing ${\rm Tr} \{ { \nu}^\alpha_i { \xi}_j^\alpha \} = C \, {\rm Tr} \, \delta_{ij}$ one has 
$\cos\theta = C d$ thus 
$ { \nu}^\alpha_i { \xi}_j^\alpha  = \left(\delta_{ij} / d \right) \cos\theta $.
For a one-component fluid in which its constituents exerts these active torques as depicted in Fig.~\ref{fig:dipole}(b), the equation of motion for ${\bm \ell}$ is (see Eq.(\ref{l3})), 
\begin{eqnarray}
\label{e15b}
\dot{\ell}_i(\vecr) + \nabla_j\left(\ell_i v_j\right) =  \varepsilon_{ijk}\ell_k\omega_j - \Gamma \left(\Omega_i - \omega_i\right) +\tilde\tau\nabla_i \rho \, ,
\end{eqnarray}
where $\tilde\tau \equiv T  a  \left(\cos\theta\right) / (md)$, with $m$ being the particles' mass.

Such active torque density generally leads to inhomogeneous stationary odd viscosity  as we now show. 
First, we expand the density $\rho = \rho_0 + \Drho$ with $\rho_0$ assumed to be constant in time (but not in space).
Then, the continuity equation becomes:
\begin{eqnarray}
\label{e15c}
\nonumber & &\partial_t\delta{\rho} + {\bf v}\cdot\nabla \rho_0 + \rho_0 \nabla \cdot {\bf v} = 0 \, .
\end{eqnarray}
Here we assumed as in previous sections that ${\bm\nabla}\vecv$ is of the order of $\Drho$.
Writing as before ${\bm \Omega} \simeq {\bm \Omega}^0 + \delta{\bm \Omega}$, 
gives (in the long-wavelength low-frequency limit) $\Omega^0_i = \tilde\tau\nabla_i\rho_0/\Gamma$, which is generally inhomogeneous but it is constant in time.
Hence, the odd viscosity is not necessarily constant (spatially), but the `odd' Navier-Stokes equation (Eq.~(6) of the main text) is still valid.

\section{Mode structure - `odd' mechanical waves}

In this section we investigate the mode structure of the odd Navier-Stokes equation.
We start by writing $[\rho,{\bf v}] \simeq [\rho_0,0] + [\Drho,{\bf v}]$ and linearize Eq.~(8) of the main text 
and the continuity equation, $\dot{\rho} + \nabla \cdot (\rho \vecv) = 0$.
Then, we perform a Fourier-Transform $[{\bf v},\Drho] = \int \frac{\D \bf k}{(2\pi)^d}[\tilde{\bf v},\delta\tilde\rho] \e^{i\left({\bf k}\cdot\vecr - s t\right)}$, which results in
\begin{eqnarray}
\label{ms1a}
\begin{pmatrix}
s + i\nu k^2  \quad 	&	-i\tilde\ell kk_\parallel  \quad		&       2i\tilde\ell kk_\perp 	\quad		&  0\\
i\tilde\ell k k_\parallel  \quad		 &	 s + i\nu k^2	  \quad		&	0 \quad		& 0 \\
-2 i\tilde\ell kk_\perp \quad 	&	0 \quad		&		   s + i\nu_L k^2  \quad		&  -kc \\
0 \quad 	&	0 \quad		&		  -kc  \quad		&  s
\end{pmatrix} 
\begin{pmatrix}
\tilde{v}_1	\\
\tilde{v}_2 \\
\tilde{v}_L \\
\tilde{h}
\end{pmatrix} = 0 \, .
\end{eqnarray}
Here a set of orthonormal unit vectors was introduced: 
$\hat{\bf k} = (k_x,k_y,k_z)/k$, $\hat{\bf e}_1 = (-k_y,k_x,0)/k_\perp$, and $\hat{\bf e}_2 = (-k_xk_z,-k_yk_z,k_\perp^2)/(kk_\perp)$,
where $k = \sqrt{{\bf k}^2}$, $k_\perp = \sqrt{k_x^2 + k_y^2}$, and $k_\parallel = {\bf k} \cdot \vecl / \ell$.
We have defined $v_{1,2} \equiv \vecv \cdot \hat{\bf e}_{1,2}$, $v_L \equiv \vecv \cdot \hat{\bf k}$, $h = c\Drho/\rho_0$,
$\nu \equiv \eta/\rho_0$, $\nu_L = 2\nu + \zeta/\rho_0$,  $\tilde\ell \equiv \ell/(4\rho_0)$, and $\vecl = \ell \hat{z}$.
Note that for vanishing viscosities ($\nu=\nu_L=0$) the matrix in Eq.~(\ref{ms1a}) is Hermitian, thus it has real  eigenvalues and the corresponding eigenvectors are orthogonal.
The fact that all eigenvalues are real means that there are no instabilities and that waves always propagate (when the viscosity is negligible).

Equation~(\ref{ms1a}) can be written in a more familiar form by using the continuity equation to write the equations for the velocity alone: 
\begin{eqnarray}
\label{ms1}
\begin{pmatrix}
	s + i\nu k^2  \quad 	&	-i\tilde\ell kk_\parallel  \quad		&	2i\tilde\ell kk_\perp 	\\
 i\tilde\ell k k_\parallel  \quad		 &	 s + i\nu k^2	  \quad		&	0 \\
-2 i\tilde\ell kk_\perp \quad 	&	0 \quad		&		  s + i\nu_L k^2 - c^2 k^2 / s   
\end{pmatrix} 
\begin{pmatrix}
\tilde{v}_1	\\
\tilde{v}_2 \\
\tilde{v}_L
\end{pmatrix} = 0 \, .
\end{eqnarray}

The eigenvalues are found using $\det{\bf M} = 0$, with ${\bf M}$ being the matrix in Eq.~(\ref{ms1a}) or(\ref{ms1}).
This results in a quartic equation
\begin{eqnarray}
\label{ms2}
\bigg[    s^2 -  c^2 k^2 + is\nu_Lk^2  \bigg] \bigg[  \left( s + i\nu k^2 \right)^2 - k^2k^2_\parallel\tilde\ell^2   \bigg]
-4s k^2k^2_\perp \tilde\ell^2 \left(	s + i\nu k^2 \right) = 0 \, . \
\end{eqnarray}
Although analytical solution is present for such an equation it does not have a lot of meaning.
Let us examine some simplifying limits.
First, inspection of Eq.~(\ref{ms1}) shows that $k_\perp$ is responsible to the coupling between longitudinal and transverse modes.
Taking $k_\perp=0$ gives 
\begin{eqnarray}
\label{ms3}
& &s_L = \frac{k}{2}\left[ -i\nu_Lk \pm  \sqrt{4c^2 - \nu_L^2k^2  }    \right] \, , \\
\label{ms4}
& &s_T = -i \nu k^2 \pm kk_\parallel\tilde\ell  \, .
\end{eqnarray}
The four modes (two longitudinal, Eq.~(\ref{ms3}), and two transverse, Eq.~(\ref{ms4})) are of decaying waves.
Unlike regular fluids, here there are two transverse propagating modes (instead of diffusive modes).
%
Another simple limit is obtained for  $k_\parallel=0$ and $\nu=0$, leading to a quadratic equation with solutions:
%
\begin{eqnarray}
\label{c10}
s = \frac{k}{2}\left[  -i\nu_L k \pm \sqrt{-\nu_L^2k^2 + 4\left( c^2 + 4 k_\perp^2\tilde\ell^2 \right)} \right] \, .
\end{eqnarray}
These solutions corresponds to two propagating (and decaying) longitudinal waves.
There are no transverse waves, neither propagating nor diffusive, in this limit.
In this case waves propagate if the $4\left( c^2 + 4 k_\perp^2\tilde\ell^2 \right)>\nu_L^2k^2$, 
revealing that when $k_\perp$ is present it can promote propagation of longitudinal waves even
when $\nu_L$ is large enough such that for normal fluids $(\ell=0)$ these waves are diffusive.

With view of the decoupled mode frequencies in Eqs.~(\ref{ms3})-(\ref{ms4}), in the long-wavelength limit the last term of Eq.~(\ref{ms2}) can be treated as perturbation 
(for $s_L$ it is $\sim k^6$ and for $s_T$ it is $\sim k^8$).
To find the correction to the longitudinal mode, $s_L$, in the long-wavelength limit we write  
\begin{eqnarray}
\label{ms5}
s^2 -  c^2 k^2 + is\nu_Lk^2 = L  \, ,
\end{eqnarray}
where $L\ll 1$.
Then, substituting $s \simeq \pm kc$ (the long wavelength solution of $s_L$ from Eq.~(\ref{ms3})) in Eq.~(\ref{ms2}) we find that $L=4\tilde\ell^2k^2k_\perp^2$.
One can now use $L$ and solve Eq.~(\ref{ms5}), yielding
\begin{eqnarray}
\label{ms6}
s_L  \simeq \pm kc \left[ 1 + \frac{16\tilde\ell^2k_\perp^2 - \nu_L^2 k^2 }{8c^2} \right] - \half i \nu_L k^2 \, .
\end{eqnarray}
The correction to the transverse modes can be found by writing
\begin{eqnarray}
\label{ms7}
\left( s + i\nu k^2 \right)^2 - k^2k^2_\parallel\tilde\ell^2  = T \, ,
\end{eqnarray}
with $T\ll 1$.
Next, we substitute the long-wavelength solution for $s_T$ (which is the same as Eq.~(\ref{ms4}))  in Eq.~(\ref{ms2}) and obtain 
$T = -\left(4\tilde\ell^3k^2k_\perp^2k_\parallel/c^2\right)\left( \tilde\ell k_\parallel \mp i \nu k  \right)$.
Using the solution for $T$ in Eq.~(\ref{ms7}) gives
\begin{eqnarray}
\label{ms8}
s_T  \simeq \left(- i \nu k^2 \pm k k_\parallel \tilde\ell\right) \left( 1 - \frac{2\tilde\ell^2k_\perp^2}{c^2} \right) \, .
\end{eqnarray}
The eigenvectors [$\vec{v}=(v_1,v_2,v_L,h)$] associated with the longitudinal and transverse waves (Eqs.~(\ref{ms6}) and (\ref{ms8})),
to lowest order in $g= 2 \tilde\ell k_{\perp}/c$ and $g_\nu = k \nu_L/c$ are:
\begin{equation}
\vec{v}_L  = (\mp i g, 0,1,\pm 1 + ig_\nu/2) \, , \qquad \vec{v}_T  = (1, \mp i, 0, - i g ) \, ,
\label{eq:vT}
\end{equation}
where $g$ is a dimensionless measure for the strength of the coupling between longitudinal and transverse modes.

The incompressible limit is formally obtained from Eq.~(\ref{ms2}) by taking $kc/s\to\infty$, which also corresponds to $g,g_\nu \ll 1$.
A predominantly transverse mode with nonzero $v_1$ satisfies  $h= - i g v_1$, or equivalently $\Drho/\rho = -i 2 \tilde\ell k_{\perp} v_1/ c^2$.  
Re-expressed in coordinate space, this gives $\Drho= - \vecl\cdot {\bm \omega}/c^2$ as also found in the main text using a different route.

\begin{figure}
	\begin{centering}
		\includegraphics[width=1\textwidth]{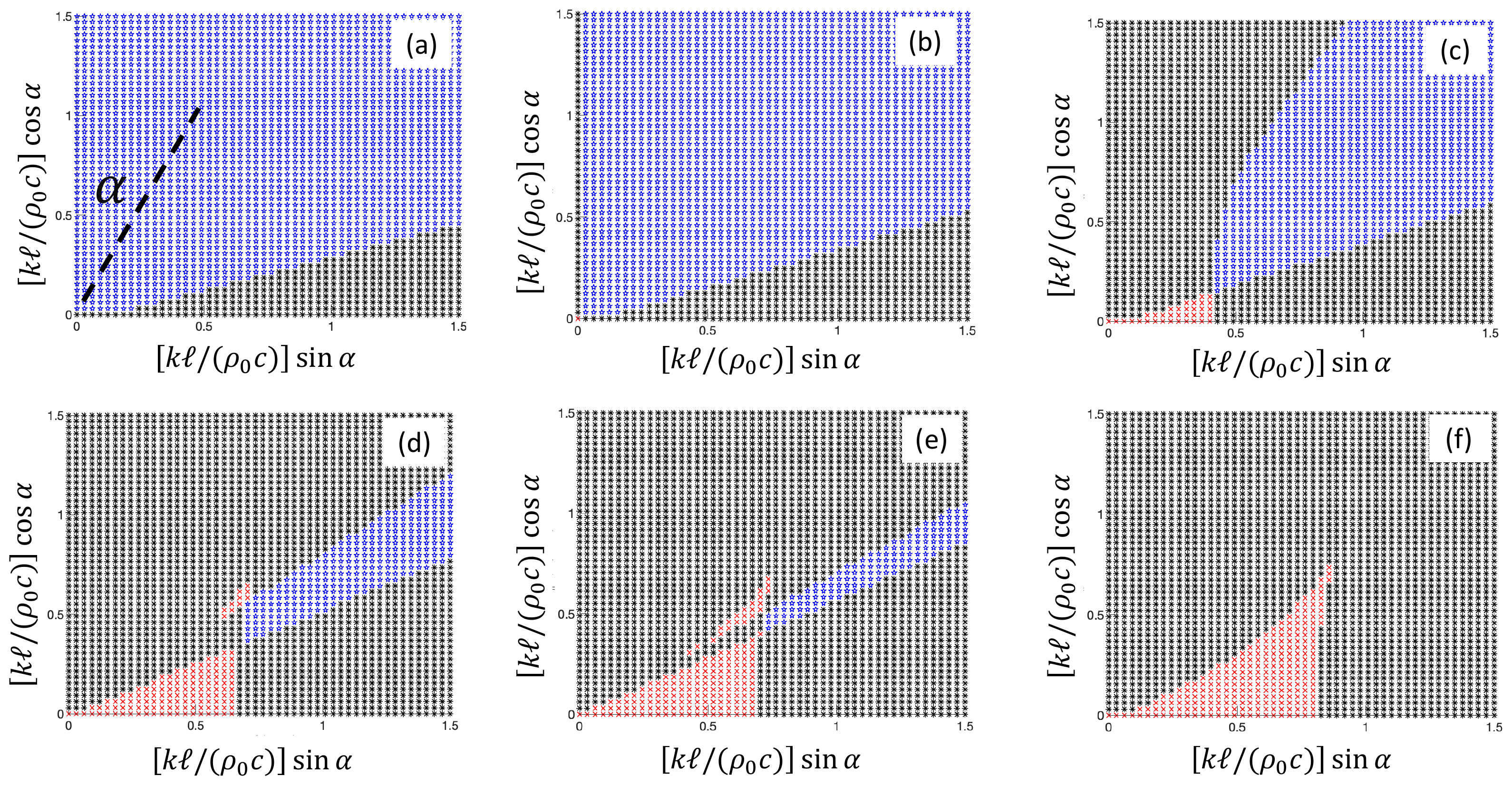}
		\par\end{centering}
	\caption{ Phase diagram in polar coordinates ($k_\perp / k = \sin\alpha$ and $k_\parallel / k = \cos\alpha$)
		for propagation of waves in the space spanned by $\vecl$ and $\bf k$, using the complete quartic equation, Eq.~(\ref{ms2}).
		In all subfigures we use constant dimensionless viscosities $\nu k / c$ and $\nu_L k / c$, 
		thus, this figure should be thought of as being plotted for finite wavenumber.
		Blue star, black asterisk, and red `x'  represent four, two, and zero propagating modes, respectively.
		In all subfigures $\nu k / c = 0.3$, while $\nu_L k / c$ is $0.9$ in (a), $2$ in (b), $2.7$ in (c), $3.5$ in (d), $3.6$ in (e) and $4$ in (f).
	}
	\label{fig:phase} 
\end{figure}

Another instructive limit is the inviscid case ($\nu,\nu_L=0$) for which Eq.~(\ref{ms2}) reduces to a quadratic equation for $s^2$,
\begin{eqnarray}
\label{c8}
& &s^4 - k^2s^2\left[ c^2  + \tilde\ell^2\left(4k_\perp^2  + k_\parallel^2 \right) \right] + k^4  k_\parallel^2 c^2\tilde\ell^2 = 0 \, .
\end{eqnarray}
Solving Eq.~(\ref{c8}) one finds
\begin{eqnarray}
\label{c9}
& &s^2 = \frac{k^2}{2}\left( c^2 + \tilde\ell^2\left(4k_\perp^2  + k_\parallel^2 \right) \pm 
\sqrt{ \left[  c^2 + \tilde\ell^2\left(4k_\perp^2  + k^2_\parallel\right) \right]^2 -  4c^2k_\parallel^2\tilde\ell^2 }  \,\right) \, .
\end{eqnarray}
The discriminant $D \equiv \left[  c^2 + \tilde\ell^2\left(4k_\perp^2  + k^2_\parallel\right) \right]^2 - 4c^2k_\parallel^2\tilde\ell^2$ 
can be shown to be strictly positive, hence, $s^2$ is real and Eq.~(\ref{c9}) also implies that $s^2>0$.
Therefore, there are four propagating modes as expected due to the fact that the matrix in Eq.~(\ref{ms1a}) is Hermitian for an inviscid odd fluid  
(see discussion after Eq.~(\ref{ms1a})).
Both solutions are admixture of the longitudinal and transverse modes.

With all the information on the limiting cases we are prepared to discuss the general mode structure.
In Fig.~\ref{fig:phase} we plot the number of propagating waves (number of solution with non-vanishing real part)
in the space spanned by $\vecl$ and $\bf k$ for constant dimensionless viscosities.
These plots are representative for $\nu k/c<1$ with various values of $\nu_L k/c$. 
When $\nu \geq 1$ only subfigure (f) is accessible because  $\nu_L>2\nu\geq2$.
Clearly $\alpha=0$ corresponds to Eqs.~(\ref{ms3})-(\ref{ms4}) such that for any nonzero value of $\ell$ transverse waves are propagating.
However, for $\nu_L k/c\geq 2$ longitudinal waves are diffusive, explaining the black vertical stripe at $\alpha=0$ in subfigure (b).
On the contrary when $\alpha=\pi/2$, Eq.~(\ref{c10}) shows that there are no transverse waves and that longitudinal waves always propagate for sufficiently large values of $\ell$,
explaining the transitions from red `x' region to black asterisk region.
For any other value of $\alpha$ all modes are a mixture of longitudinal and transverse modes.
The blue star region can be thought of as an effectively inviscid region in the sense that all four modes propagate in this region 
(of course these viscosities results in decaying  waves) and for large enough values of $\ell$, Eq.~(\ref{c9}) indeed gives the frequencies for the four modes.
%
Note that the long wavelength limit corresponds to the region around the origin in subfigure (a) alone.
In all other subfigures $\nu_L k^2 > kc$ such that they must be thought of as being plotted for finite wavenumber.
It is interesting that the red `x' region does not need to be continuous, showing the complicated dependence of the eigenfrequencies on $\alpha$.

\section{Breakdown of Kelvin's circulation theorem}
The strength of the vortex tube, also referred to as circulation, is defined as
\begin{eqnarray}
\label{h1}
\Gamma = \oint_c \vecv \cdot \D {\bf s} = \int_A 2{\bm \omega} \cdot \D {\bf A} \, ,
\end{eqnarray}
where $A$ is the area bound by the closed contour $c$ and ${\bm \omega} = \half\nabla\times\vecv$. Kelvin's circulation theorem (and also Helmholtz's first theorem) states that the circulation strength does not change in time, $\D \Gamma / \D t =0$. 

Let us consider the simple case of an inviscid ($\eta=0$) incompressible (constant density) `odd' fluid. Taking the curl of the odd Navier-Stokes equation, Eq.~(12) of the main text, while using $\nabla\times\nabla\times\vecv = \nabla\nabla\cdot\vecv - \nabla^2\vecv = -\nabla^2\vecv$, yields the following `odd' vorticity equation
\begin{eqnarray}
\label{h2}
\frac{ D \omega_i}{D t}  = \frac{1}{2\rho} \varepsilon_{ijk} \left( \vecl\cdot\nabla \right) \nabla_j \omega_k + {\bm \omega} \cdot \nabla v_i \, .
\end{eqnarray}
The first term is the novel odd term that only appears in 3D (in 2D $\vecl\cdot\nabla X = 0$ for any $X$).  One can then write the vortex strength as,
\begin{eqnarray}
\label{h3}
\frac{\D \Gamma}{\D t} = 2 \int_A \frac{ D {\bm \omega}}{D t} \cdot \D {\bf A}  = 
 \int_A \left[  \varepsilon_{ijk} \hat{n}_i \left( \vecl\cdot\nabla \right) \nabla_j \omega_k + 2 {\bm \omega} \cdot \nabla v_i \hat{n}_i \right] \D A  \, ,
\end{eqnarray}
with $ \hat{\bf n}$ being the surface direction. Integrating by parts the second term and using the divergence theorem gives
\begin{eqnarray}
\label{h4}
\frac{\D \Gamma}{\D t} =  \int_A \left( \vecl\cdot\nabla \right) \left(\nabla\times{\bm\omega}\right) \cdot \D{\bf A} \, ,
\end{eqnarray}
which does not vanish, such that the vortex strength is not conserved in 3D odd fluids. Because odd viscosity originates in body torques that generally results in non-conservative forces, for which Kelvin's circulation theory does not hold, one might expect such breakdown for odd fluids. However, the case considered here is of constant $\vecl$, which is a result of a constant torque that does not produce a non-conservative force. It is for this reason that Kelvin's circulation theory still holds for 2D odd fluids.